\documentclass{osa-article}

\usepackage{bm}
\usepackage{cancel}
\usepackage{booktabs} % To thicken table lines

\journal{osajournal}
\articletype{Research Article}
\newcommand{\ket}[1]{\bigl| #1 \bigr>} % for Dirac bras
\newcommand{\bra}[1]{\bigl< #1 \bigr|} % for Dirac kets
\newcommand{\braket}[2]{\bigl< #1 \vphantom{#2} \bigr|
 \bigl. #2 \vphantom{#1} \bigr>} % for Dirac brackets
\newcommand{\op}[1]{\hat{#1}}
\newcommand{\vop}[1]{\hat{\vec{#1}}}
\newcommand{\uv}[1]{\ensuremath{\hat{\mathbf{#1}}}} % for unit vector
\renewcommand{\vec}[1]{{\mathbf{\bm{#1}}}}
\newcommand{\abs}[1]{\left| #1 \right|} % for absolute value
\newcommand{\norm}[1]{\left\lVert#1\right\rVert}
\renewcommand{\Im}{\text{Im}}
\newcommand{\tr}{\text{Tr}}

\newcommand{\berryc}{\mathcal{A}}
\newcommand{\uc}{{\text{cell}}} % unit cell
\newcommand{\crys}{{\text{crys}}} % unit cell
\newcommand{\vuc}{V_{\text{cell}}} % volume unit cell
\newcommand{\vcrys}{V_{\text{crys}}} % volume crystal

\newcommand{\hBN}{hBN}

\newcommand{\rand}[1]{\breve{#1}}
\newcommand{\pt}[1]{\bar{#1}}
\newcommand{\tpt}[1]{{#1}}
\newcommand{\vg}{{VG}}
\renewcommand{\lg}{{LG}}
\newcommand{\tra}{{\text{tra}}}
\newcommand{\ter}{{\text{ter}}}
\newcommand{\anom}{{\text{anom}}}
\newcommand{\mix}{{\text{mix}}}
\newcommand{\ntra}{{\text{nontra}}}

\newcommand{\pol}{\mu}
\newcommand{\com}{\vec{\mathcal{D}}}
\newcommand{\ccom}{\vec{\mathcal{Q}}}
\newcommand{\sad}[1]{{\bar{#1}}} % saddle point indicator

\begin{document}

\title{Introduction to theory of high-harmonic generation in solids: tutorial}

\author{Lun Yue\authormark{1,3} and Mette B. Gaarde\authormark{1,4}}
\address{\authormark{1}Department of Physics and Astronomy, Louisiana State University, Baton Rouge, Louisiana 70803-4001, USA\\
\email{\authormark{3}lun\_yue@msn.com}
\email{\authormark{4}mgaarde1@lsu.edu}}
% \email{\authormark{*}opex@osa.org} %% email address is required


\begin{abstract}
  High-harmonic generation (HHG) in solids has emerged in recent years as a rapidly expanding and interdisciplinary field, attracting attention from both the condensed-matter and the atomic, molecular, and optics communities. It has exciting prospects for the engineering of new light sources and the probing of ultrafast carrier dynamics in solids, and the theoretical understanding of this process is of fundamental importance.
  This tutorial provides a hands-on introduction to the theoretical description of the strong-field laser-matter interactions in a condensed-phase system that give rise to HHG. 
  We provide an overview ranging from a detailed description of different approaches to calculating the microscopic dynamics and how these are intricately connected to the description of the crystal structure, through 
  the conceptual understanding of HHG in solids as supported by the semiclassical recollision model, and finally a brief description of how to calculate the macroscopic response. 
  We also give a general introduction to the Berry phase, and we discuss important subtleties in the modelling of HHG, such as the choice of structure and laser gauges, and the construction of a smooth and periodic structure gauge for both nondegenerate and degenerate bands.  The advantages and drawback of different structure and laser-gauge choices are discussed, both in terms of their ability to address specific questions and in terms of their numerical feasibility.
\end{abstract}

\section{Introduction} \label{sec:intro}

High-harmonic generation (HHG) is a extremely nonlinear optical process where a macroscopic system irradiated by  intense laser light emits coherent radiation with frequencies many times that of the driving laser field. HHG in gases has facilitated the generation of attosecond light pulses, the probing and control of electrons on their natural timescales, and more generally 
laid the foundation for attosecond science \cite{Corkum2007review, Krausz2009review}. The last decade has seen HHG extended to  systems in the condensed phase. Since the initial observation of nonperturbative HHG\footnote{Earlier measurements of HHG in reflection \cite{Burnett1977APL, Carman1981PRL, Linde1995PRA, Norreys1996PRL} and transmission geometries \cite{Chin2001PRL} have not explicitly established the non-pertubative nature of HHG.} by Ghimire et al. \cite{Ghimire2011NPhys} in a bulk solid in 2011, HHG has been observed in semiconductors \cite{Schubert2014NPhoton, Vampa2015Nature, You2017NPhys, Jiang2019JPB}, dielectrics \cite{Luu2015Nature, Garg2018NPhoton}, rare-gas solids \cite{Ndabashimiye2016Nature}, monolayer materials \cite{Yoshikawa2017Science, Liu2017NPhys, Hafez2018Nature, Yoshikawa2019NCommun}, nanostructures \cite{Han2016NCommun, Vampa2017NPhys, Sivis2017Science}, amorphous solids \cite{You2017NCommun}, doped systems \cite{Nefedova2021APL} and topological insulators \cite{Bai2021NPhys, Schmid2021Nature}. Solid-state HHG has exciting prospects for new compact attosecond light-source technologies \cite{Luu2015Nature, Han2016NCommun, Vampa2017NPhys, Sivis2017Science, Garg2018NPhoton}, as well as novel ultrafast spectroscopy methods capable of probing band structures \cite{Vampa2015PRL, Uzan2020NPhoton}, Berry curvatures \cite{Luu2018NCommun, Liu2017NPhys}, and topological effects \cite{Chacon2020PRB, Bai2021NPhys, Schmid2021Nature}.

\begin{figure}
  \centering
  \includegraphics[width=1.0\textwidth, clip, trim=0 0cm 0 0cm]{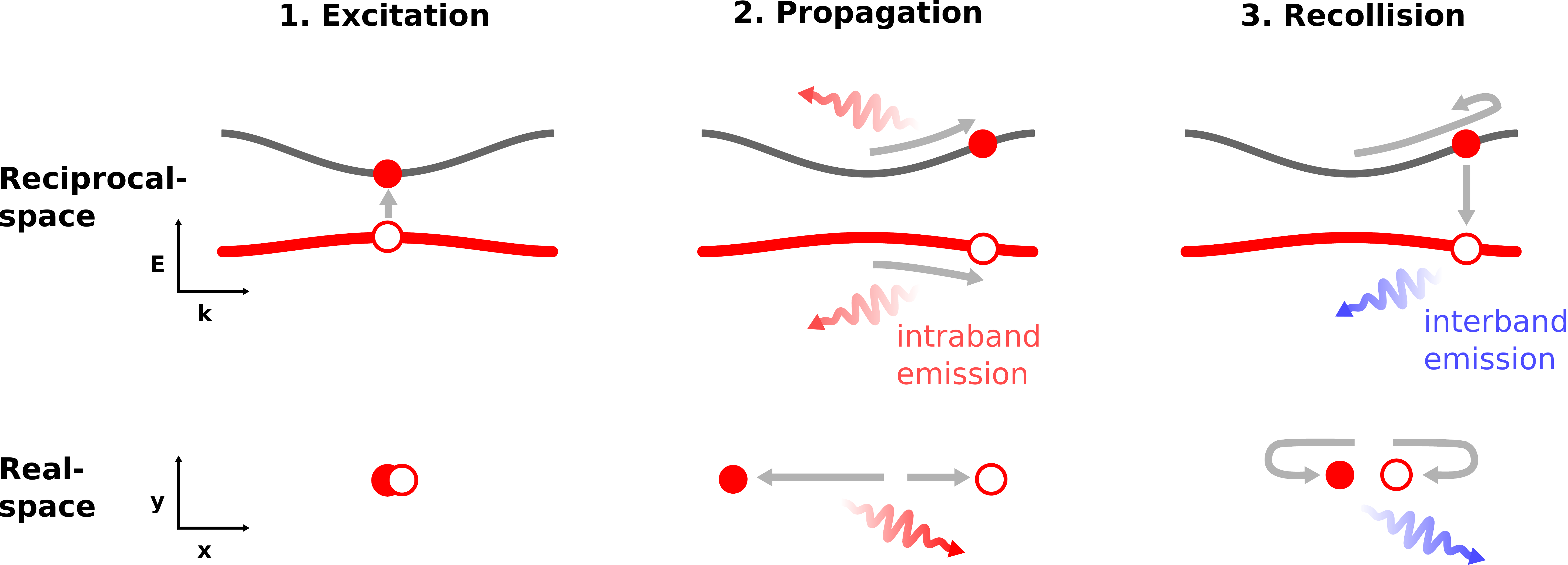}
  \caption{Sketch of the recollision model for HHG. Upper panels: tunneling, propagation and recombination in reciprocal space. Lower panels: the equivalent physics in real space.}
  \label{fig:intro_1}
\end{figure}

An intuitive semiclassical understanding of HHG in condensed phase systems as a three-step process  is illustrated in Fig.~\ref{fig:intro_1}, in both reciprocal space (top panels) and real space (bottom panels). 
In reciprocal space the first step is the creation of an electron-hole pair via excitation of an electron from the valence band to the conduction band, usually by tunneling near the minimum band gap. Per the acceleration theorem \cite{Bloch1929ZPhys, Houston1940PR}, the time-dependent crystal momentum will follow the time-dependent vector potential, and the resulting carrier motion in the (nonparabolic) bands will lead to the emission of nonperturbative {\it intraband} harmonics \cite{Ghimire2011NPhys, Ghimire2012PRA}. At the same time, also {\it interband} harmonic radiation is emitted via recombination, with frequencies corresponding to the instantaneous band gap \cite{Vampa2014PRL, Vampa2015Nature}. The interband radiation results from the coherence of the electron-hole pair, and is emitted when a stationary phase condition is satisfied for the phase that is accumulated during propagation in the bands. 
In real space, the three steps again consist of tunneling, which creates the electron-hole pair; propagation, which accelerates them apart in space and leads to intraband emission; and recollision, when the electron and hole reencounter each other in space and drive interband emission via recombination. The recollision corresponds to the stationary phase condition from the reciprocal picture being exactly satisfied. The real-space semiclassical model represents the generalization of the recollision model for gas-phase HHG  \cite{Corkum1993PRL, Lewenstein1994PRA} to the condensed phase.
Although the picture illustrated in Fig.~\ref{fig:intro_1} is based on a semiclassical understanding of the HHG process, its interpretation 
has been supported by time-dependent strong-field light-matter simulations such as the single-particle time-dependent Schr\"odinger equation (TDSE) \cite{Wu2015PRA, Guan2016PRA, Ikemachi2017PRA, Li2019PRL}, time-dependent density-functional theory \cite{Runge1984PRL, Tancogne-Dejean2017PRL, Bauer2018PRL, Yu2019PRA, Jensen2021PRB}, SBEs \cite{Golde2008PRB, Haug2004book, Kira2012book} and density matrix approaches \cite{Vampa2014PRL, Floss2018PRA, Yue2020PRA}. 

Any calculation of HHG from a condensed-phase material interacting with a strong laser field involves at least three different types of calculations: Finding the initial state of the material, solving an equation of motion (EOM) in order to describe the time-evolution, and calculating the observable current(s) that will make up the harmonic spectrum. For a periodic material, these calculations are most conveniently performed in reciprocal space, using Bloch states, so that both the structure and the dynamics of the system are described in terms of its band structure. As we will describe in more detail in the next couple of paragraphs, these calculations then take the form of: (i) Calculating the band structure, including relevant matrix elements of the momentum or position operators, by solving the time-independent Schr\"odinger equation (TISE). (ii) Solving the EOM after choosing a
{\it structure} as well as a {\it laser} gauge for calculating the dynamics, where the gauge choice will determine the exact form of the EOM. (iii) Calculating the total current and perhaps different contributions to the current that have different physical meaning, such as the intraband and interband currents described above, or the anomalous current driven by the Berry curvature \cite{Luu2018NCommun, Liu2017NPhys}.  

Starting with the structure calculation,  there is an important, and sometimes subtle, complication that deserves special attention in the theoretical description of carrier dynamics in a solid. Since the different crystal momenta $\vec{k}$ are treated independently in the structure calculation in reciprocal space, there is a $\vec{k}$-dependent phase arbitrariness in the Bloch states, i.e. a gauge freedom, which we will henceforth  refer to as the \textit{structure gauge}. It is often favorable to pick a gauge (generally referred to as gauge fixing) where the Bloch functions are smooth, especially if one considers the dynamics in terms of electrons and holes explicitly moving along the bands. In the case of large carrier excursions in the Brillouin zone (BZ), as is  often the case for carrier motion beyond the perturbative regime and the breakdown of the notion of effective masses, a BZ-periodic gauge is also required. The construction of such a smooth and BZ-periodic structure gauge is intricate, especially in the case of bands containing degeneracies \cite{Resta1994review, Vanderbilt2018book, Marzari2012review}. As we will discuss in more detail in Sec.~\ref{sec:berry}, the structure gauge is closely related to the Berry phase \cite{Berry1984PRSL} that is accumulated under adiabatic motion, which in condensed matter theory has had profound implications such as the development of modern theory of polarization \cite{KingSmith1993PRB} and the discovery of the quantization of adiabatic transport \cite{Thouless1983PRB}.

In the description of the time-dependent interaction between the crystal and the strong laser field there is another gauge freedom, namely the choice of {\it laser gauge}. While all physical observables in principle are gauge-invariant in a complete basis, for computational purposes  the number of bands and number of $\vec{k}$-points in the BZ should be truncated. This basis truncation depend critically on the choice of laser gauge \cite{Aversa1995PRB, Foldi2017PRB, Virk2007PRB, Taghizadeh2018, Yue2020PRA}. In addition, the choice of laser gauge is linked to the fixing of the structure gauge \cite{Yue2020PRA}, since the band-coupling terms are different in different laser gauges \cite{Blount1962book}. 

Finally, while the microscopic theory of HHG can be solved in the framework of the dipole approximation, the experimentally measurable signal is actually the macroscopic HHG response. Hence, a realistic treatment of HHG should involve the effect of beam propagation in the medium, as well as the propagation of the radiation from the near-field at the sample to the far-field at the detector. These effects requires the solution of the coupled microscopic response to Maxwell's equations, which is very computationally demanding \cite{Brabec2000review, Kolesik2002PRL, Boyd2008book, Gaarde2011PRA, Yabana2012PRB, Floss2018PRA}.

This tutorial provides a hands-on introduction to doing calculations of strong-field laser-matter interactions in the condensed phase, with an emphasis on the generation of high-order harmonics. Although the tutorial is primarily aimed at scientists specifically interested in performing calculations of ultrafast condensed-phase dynamics, we believe that many of the concepts and descriptions will be useful to anyone wishing to gain a deeper understanding of the current state and capabilities of solid-state-HHG theory. Indeed, the decomposition of the current, the semiclassical recollision model, and the macroscopic effects can directly be used to interpret experiments. It is also worth noting that HHG in solids is an inherently interdisciplinary scientific field, attracting researchers from both the condensed-matter physics community and the strong-field / attoscience community which originated with atomic, molecular, and optical physicists. We thus also aim to provide some unification of concepts across this diverse community of scientists.

We start the tutorial in Sec.~\ref{sec:berry} with an introduction to adiabatic states and related concepts such as Berry phase, connections and curvatures, all of which will be useful for reading the later sections. Section~\ref{sec:ham} deals with the time-independent problem of a crystalline solid, and provides methods for the construction of a smooth and BZ-periodic structure gauge. The random gauge, the parallel transport (PT) gauge, the twisted PT (TPT) gauge, the Wannier gauge, as well as degenerate bands will be covered. Section~\ref{sec:tdse} treats the time-dependent microscopic problem of HHG, and presents the relevant EOMs in the velocity gauge (VG) and the length gauge (LG), as well as in the time-dependent adiabatic Houston basis. The advantages and drawbacks of the different methods are compared, and we present a concrete calculation example for HHG in a monolayer material. Section~\ref{sec:saddle} is about the saddle-point method and the semiclassical solutions to the saddle point equations, i.e. the recollision model. Sec.~\ref{sec:macro} gives a brief introduction to the macroscopic propagation schemes for HHG and provides an example of a near-field to far-field propagation scheme and discusses the spatio-spectral properties of the far-field spectrum.

Atomic units, where the reduced Planck constant, the elementary charge, the Bohr radius and the electron mass are set to unity, are used throughout this work unless indicated otherwise.

\section{Adiabatic states, Berry connections and curvatures} \label{sec:berry}
We start this tutorial by briefly introducing the adiabatic states and some general concepts that could be helpful for the understanding of the rest of the tutorial. Concepts such as Berry phases, Berry connections, Berry curvatures and gauge fixing will be discussed. 
For further reading, see e.g. Refs.~\cite{Berry1984PRSL, Resta1994review, Sakurai1994book, Xiao2010review, Vanderbilt2018book, Madsen2021review}. 

\subsection{The adiabatic states} \label{sec:berry_adiab}
When describing laser-matter interactions, the Hamiltonian $\op{H}(t)$ is generally time-dependent, and it is often useful to describe the dynamics of the system using the {\it adiabatic states}. These are defined as the eigenstates of the instantaneous Hamiltonian at time $t$
\begin{equation}
  \label{eq:adiab_1}
  \op{H}(t) \ket{n(t)} = \epsilon_n(t) \ket{n(t)},
\end{equation}
with $\ket{n(t)}$ an adiabatic state and $\epsilon_n(t)$ its energy.
The TDSE reads
\begin{equation}
  \label{eq:adiab_2}
  i \ket{\dot\Psi(t)} = \op{H}(t) \ket{\Psi(t)},
\end{equation}
with $\ket{\Psi(t)}$ the quantum state, and the diacritic dot is used henceforth to denote full time-derivatives. It is important to note that an instantaneous eigenstate satisfying Eq.~\eqref{eq:adiab_2} at a given $t$ is not uniquely defined, but rather carries an arbitrary phase factor, i.e. there is a \textit{gauge freedom}. This gauge freedom and its consequences will recur many times throughout this document. 

We consider the case where the adiabatic states are nondegenerate\footnote{The degenerate case will be discussed for the Brillouin problem in Sec.~\ref{sec:ham_deg}.} for all $t$. The wave function expanded into the adiabatic states reads
\begin{equation}
  \label{eq:adiab_3}
  \ket{\Psi(t)} = \sum_nc_n(t) e^{i\theta_n(t)} \ket{n(t)},
\end{equation}
with $\theta_n(t) = -\int^t\epsilon_n(t')dt'$ the dynamical phase. We can get the EOM for the time-dependent coefficients $c_n(t)$ by inserting Eq.~\eqref{eq:adiab_3} into the TDSE of Eq.~\eqref{eq:adiab_2} and projecting onto $\bra{n(t)}$,
\begin{equation}
  \label{eq:adiab_4}
  \dot{c}_n(t) =  - \braket{n(t)}{\dot{n}(t)} c_n(t)
  - \sum_{m\ne n} c_m(t) \braket{n(t)}{\dot{m}(t)} e^{i[\theta_m(t)-\theta_n(t)]}.
\end{equation}
The first term on the right-hand side of Eq.~\eqref{eq:adiab_4} depends only on the $n$th state, while the second term involves the nonadiabatic couplings $\braket{n(t)}{\dot{m}(t)}$ that is responsible for the transfer of population between adiabatic states.
Rewriting $\braket{n(t)}{\dot{m}(t)} = -\bra{n(t)}\dot{\op{H}}(t)\ket{m(t)}/ \left[\epsilon_n(t)-\epsilon_m(t)\right]$ for $m\ne n$, one can see that the nonadiabatic couplings are generally large when the adiabatic states are close to each other in energy. 

In the adiabatic approximation, the  terms in Eq.~\eqref{eq:adiab_4} involving the nonadiabatic couplings are neglected. For a system that start in the $n$th adiabatic state, $\Psi(t_0) = \ket{n(t_0)}$, such that $c_n(t_0)=\delta_{mn}$, the time evolution proceeds as: 
\begin{equation}
  \label{eq:adiab_6}
  \ket{\Psi(t)} = e^{i\theta_n(t)} e^{i\gamma_n(t)} \ket{n(t)},
\end{equation}
where
\begin{equation}
  \label{eq:adiab_7}
  \gamma_n(t) = i \int_{t_0}^t dt' \braket{n(t')}{\dot{n}(t')}
\end{equation}
is the Berry phase or the geometric phase \cite{Berry1984PRSL, Resta1994review}.

\subsection{The Berry phase, connection and curvature}
The Berry phase $\gamma_n(t)$ in Eq (7) emerged  in the adiabatic state basis as the additional phase accumulated by a state beyond its energy evolution. The Berry phase is often termed the geometrical phase because of its geometric properties. These can be realized by assuming without loss of generality that the Hamiltonian depends on time through a set of parameters $\vec{R}(t)=[R_1(t),R_2(t),\dots]$, such that $\op{H}(t)=\op{H}[\vec{R}(t)]$. The Berry phase can now be rewritten
as an integral over a path $\mathcal{C}$ through this parameter space, from $\vec{R}(t_0)$ to $\vec{R}(t)$:
\begin{equation}
  \label{eq:adiab_8}
  % \gamma_n(t) = i \int_{C} \bra{n(\vec{R})}\partial_{\vec{R}}\ket{n(\vec{R})} \cdot d\vec{R},
  \gamma_n(\mathcal{C}) = \int_{\mathcal{C}} \vec{\berryc}_n(\vec{R}) \cdot d\vec{R},
\end{equation}
and the purely real Berry connection is defined as
\begin{equation}
  \label{eq:adiab_9}
  \vec{\berryc}_n(\vec{R}) \equiv i \bra{n(\vec{R})}\partial_{\vec{R}}\ket{n(\vec{R})},
\end{equation}
with $\partial_{\vec{R}}=[\partial_{R_1},\partial_{R_2},\dots]$ the operator for the partial derivatives. Note that the Berry phase in this formulation only depends on the path and no longer on the time duration.

For a closed path, the Berry phase can be rewritten to a surface integral by application of Stoke's theorem, with $\mathcal{S}(\mathcal{C})$ denoting a surface enclosed by $\mathcal{C}$. For convenience, we assume that the parameter space $\vec{R}$ is three-dimensional, such that $\partial_{\vec{R}}=\nabla_{\vec{R}}$, although derivations for higher dimensions proceed similarly. We have
\begin{equation}
  \label{eq:adiab_11}
  \gamma_n(\mathcal{C}) = \oint_{\mathcal{C}} \vec{\berryc}_n(\vec{R}) \cdot d\vec{R} = \iint_{\mathcal{S}(\mathcal{C})} \left[ \nabla_{\vec{R}} \times \vec{\berryc}_n(\vec{R})\right] \cdot d\vec{S},
\end{equation}
where the integrand in the surface integral is the Berry curvature
\begin{equation}
  \label{eq:adiab_12}
  \vec{\Omega}_n(\vec{R}) \equiv \nabla_{\vec{R}} \times \vec{\berryc}_n(\vec{R}).
\end{equation}
The Berry curvature is important because, like the Berry phase, it is independent under a gauge transform. This is in contrast to the Berry connection, which transforms as $\vec{\berryc}_n \rightarrow \vec{\berryc}_n + \nabla_{\vec{R}}\chi$ under a gauge transform of the adiabatic states $\ket{n}\rightarrow \ket{n}e^{-i\chi(\vec{R})}$, with $\chi$ an arbitrary real function. 
Note that the Berry connection and the Berry curvature transform similarly to the vector potential and magnetic field from classical electromagnetism, respectively.

The Berry curvature can be expressed in a form that is useful for numerical evaluations \cite{Gradhand2012review}. Using $\nabla\times(\psi\nabla\phi) = \nabla \psi \times \nabla\phi$, the resolution of identity, and $\braket{m(\vec{R})}{\nabla_{\vec{R}}n(\vec{R})} [\epsilon_n(\vec{R})-\epsilon_m(\vec{R})] = \bra{m(\vec{R})} \nabla_{\vec{R}}\op{H}(\vec{R})\ket{n(\vec{R})}$ for $m\ne n$ obtained by taking the gradient of Eq.~\eqref{eq:adiab_2}, the Berry curvature can be written
\begin{equation}
  \label{eq:adiab_13}
  \begin{aligned}
    \vec{\Omega}_n(\vec{R})
    & = i \bra{\nabla_{\vec{R}}n(\vec{R)}} \times \ket{\nabla_{\vec{R}}n(\vec{R)}} \\
    & = i \sum_{m}\braket{\nabla_{\vec{R}}n(\vec{R)}}{m(\vec{R})} \times \braket{m(\vec{R})}{\nabla_{\vec{R}}n(\vec{R)}} \\
    & = i \sum_{m\ne n}
    \frac{\bra{n(\vec{R)}}\nabla_{\vec{R}}\op{H}(\vec{R})\ket{m(\vec{R})}
      \times \bra{m(\vec{R})}\nabla_{\vec{R}}\op{H}(\vec{R})\ket{n(\vec{R)}}}
    {\left[\epsilon_m(\vec{R})-\epsilon_n(\vec{R})\right]^2}.
  \end{aligned}
\end{equation}
From this expression, the Berry curvature for the $n$th adiabatic state is intuitively seen to originate from the coupling to all the other adiabatic states, and the closer in energy a neighbouring adiabatic state is, the larger the contribution.
Equation~\eqref{eq:adiab_13} is useful for numerical evaluations since no derivatives are taken with respect to the adiabatic states.

As mentioned above, there is a gauge freedom in the adiabatic states, i.e. a phase-arbitrariness in the definition of the adiabatic states. One way to remove this arbitrariness of the phase is to enforce the \textit{parallel transport} (PT) gauge condition on the eigenstates
\begin{equation}
  \label{eq:adiab_10}
  0 = \braket{n(t)}{\dot{n}(t)} = \dot{\vec{R}} \cdot \bra{n(\vec{R})}\partial_{\vec{R}}\ket{n(\vec{R})}.
\end{equation}
In the PT gauge, the ``velocity'' of the state is thus perpendicular to the state itself.
With this gauge choice, the eigenstate $\ket{n(t)}$ is always single-valued as a function of $t$, and the integrand in Eq.~\eqref{eq:adiab_8} is zero along the path. However, if $\mathcal{C}$ is a closed path in parameter space (e.g. under periodic motion) such that $\vec{R}(t_0)=\vec{R}(T)$, there is no guarantee that $\ket{n[\vec{R}(t_0)]}$ is equal to $\ket{n[\vec{R}(T)]}$ under PT. The phase difference between $\ket{n[\vec{R}(t_0)]}$ and $\ket{n[\vec{R}(T)]}$ is exactly the Berry phase $\gamma_n(T)$. If $\gamma_n(T)\ne 0$, the adiabatic state $\ket{n}$ in PT gauge is thus a multivalued function of $\vec{R}$ (but still a single-valued function of $t$).
Instead of the PT gauge, a sometimes more convenient gauge is the periodic gauge, in which one applies  the additional requirement that the adiabatic state is single-valued along the path $\mathcal{C}$ in parameter space.
Later, in Sec.~\ref{sec:ham} of this tutorial we will present methods to construct such a gauge in crystalline systems.

Finally, by neglecting adiabatic coupling in the second term on the right hand side of Eq.~\eqref{eq:adiab_4}, the quantum evolution of $\ket{\Psi(t)}$ in Eq.~\eqref{eq:adiab_6} is correct only to zeroth order in $\dot{\vec{R}}$, i.e. if it starts in the $n$th adiabatic state, it stays in that state, with no transition to other adiabatic states.
The result can be extended to first order by making the ansatz $\ket{\Psi^{(1)}}=e^{i\theta_n} e^{i\gamma_n} \left[\ket{n} + \dot{\vec{R}}\ket{\delta n}\right]$ in the TDSE and solving for $\ket{\delta n}$, resulting in the state \cite{Vanderbilt2018book, Xiao2010review}
\begin{equation}
  \label{eq:adiab_14}
  \ket{\Psi^{(1)}(t)} =e^{i\theta_n(t)} e^{i\gamma_n(t)} \left[\ket{n(t)}
    + i \sum_{m\ne n} \frac{\braket{m(t)}{\dot{n}(t)}}{\epsilon_m(t)-\epsilon_n(t)}\ket{m(t)}
  \right].
\end{equation}
The second term on the right-hand side of Eq.~\eqref{eq:adiab_14} results in the so-called anomalous current, which we will return to in Sec.~\ref{sec:tdse}. This expression, together with the Berry phase, has been crucial for the development of the theory of adiabatic charge transport \cite{Thouless1983PRB} and the modern theory of polarization \cite{KingSmith1993PRB, Resta1994review}. 

\section{The time-independent problem and structure gauges} \label{sec:ham}
In this section, we discuss the time-independent problem of a crystalline solid. While for a perfect crystal it can be argued that everything boils down to the TISE and Bloch's theorem, the Bloch states can have an arbitrary crystal-momentum-dependent phase, representing a gauge degree of freedom. This section is devoted to the discussion of this gauge freedom and presents methods for gauge fixing.
We start with Bloch's theorem, the choice of crystal conventions, and how to calculate relevant matrix elements. We then discuss in Secs.~\ref{sec:ham_random}-\ref{sec:ham_wannier} four different gauge choices and conclude with Sec.~\ref{sec:ham_deg} for gauge fixing in the  nondegenerate case.

\subsection{Bloch states and coupling matrix elements}
The single-particle eigenstates satisfying Bloch's theorem are solutions to the crystalline TISE
\begin{equation}
  \label{eq:ham_1}
  \op{H}_0\phi_n^{\vec{k}}(\vec{r}) = E_n^{\vec{k}} \phi_n^{\vec{k}}(\vec{r}),
\end{equation}
where the Hamiltonian $\op{H}_0=\op{T} + \op{V}$ contains the the kinetic energy operator $\op{T}=\op{\vec{p}}^2/2$ and the periodic potential satisfying $V(\vec{r})=V(\vec{r}+\vec{R})$ with $\vec{R}$ a lattice vector. 
The band energies $E_n^{\vec{k}}$ and Bloch states $\phi_n^{\vec{k}}(\vec{r})$ in Eq.~\eqref{eq:ham_1} depend on the band index $n$ and the crystal momentum $\vec{k}$. According to Bloch's theorem, the Bloch states can be written $\phi_n^{\vec{k}}(\vec{r})=e^{i\vec{k}\cdot \vec{r}}u_n^{\vec{k}}(\vec{r})$, with $u_n^{\vec{k}}(\vec{r})$ a lattice-cell-periodic function. 
While in this tutorial we focus on systems without electron-electron interactions, it has been shown that most of the conclusions can be extended to interacting systems, provided that the interacting system is adiabatically connected to a noninteracting one\footnote{Two systems are adiabatically connected if one can go to the other by following a suitable path in parameter space - i.e. in this case that the interacting system can be reduced to the noninteracting one by an adiabatic path that takes the finite interaction strength to zero.} \cite{Vanderbilt2018book, Xiao2010review}.

Before proceeding, some details on the choice of lattice are warranted. We consider a $D$-dimensional supercell with the Bravais lattice
\begin{equation}
  \label{eq:ham_1a}
  \vec{R}=\sum_{d=1}^{D}m_d\vec{a}_d, \qquad m_d\in\left[-\frac{N_d}{2},\frac{N_d}{2}-1\right],
\end{equation}
that has volume $\vcrys=N\vuc$, where $N=\prod_{d=1}^DN_d$ is the total number of unit cells, $\vuc$ is the unit cell volume, and $D$ the lattice dimension.
To avoid surface effects,  periodic boundary conditions $\phi_n^{\vec{k}}(\vec{r}+N_d\vec{a}_d) = \phi_n^{\vec{k}}(\vec{r})$ are enforced \cite{Haug2004book}, which results in discretized crystal momenta in a BZ
\begin{equation}
  \label{eq:ham_1b}
  \vec{k}
  =\sum_{d=1}^D\frac{n_d}{N_d}\vec{b}_d
  =\sum_{d=1}^D\kappa_d\uv{b}_d
  , \qquad n_d\in\left[-\frac{N_d}{2},\frac{N_d}{2}-1\right],
\end{equation}
where we have defined the reduced reciprocal coordinates $2\kappa_d\in[-\norm{\vec{b}_d}, \norm{\vec{b}_d}]$ along the reciprocal vector directions with grid spacings $\Delta\kappa_d=\norm{\vec{b}_d}/N_d$, and $\uv{b}_d=\vec{b}_d/\norm{\vec{b}_d}$ is an unit vector.
Two useful orthogonality relations are
\begin{align}
  \sum_{\vec{k}\in BZ}e^{i\vec{k}\cdot(\vec{R}-\vec{R}')} & = N\delta_{\vec{R},\vec{R}'}, \label{eq:ham_1c}\\
  \sum_{\vec{R}}e^{-i(\vec{q}-\vec{k})\cdot \vec{R}} & = N\delta_{\vec{q},\vec{k}} \label{eq:ham_1d}, 
\end{align}
and we define the supercell and unit cell inner products as
\begin{align}
  \label{eq:ham_1e}
  \braket{f}{g}_\crys & \equiv \int_\crys f^*(\vec{r})g(\vec{r})d \vec{r} \\
  \braket{f}{g}_\uc & \equiv \int_\uc f^*(\vec{r})g(\vec{r})d \vec{r}.
\end{align}

As we will see in Section \ref{sec:tdse}, the matrix elements of interest for solving the time-dependent problem are those of the momentum and  position operators, $\op{\vec{p}}$ and $\vop{r}$, respectively. 
To provide an example of steps involved in the evaluation of matrix elements, it can be seen that operators $\op{O}(\vop{p})$ involving the momentum operator $\op{\vec{p}}$ are diagonal in the crystal momentum:
\begin{equation}
  \label{eq:ham_1f}
  \begin{aligned}
    \bra{\phi^{\vec{q}}_{m}} \op{O}(\vop{p})  \ket{\phi^{\vec{k}}_n}_\crys
    =&\int_\crys\phi^{\vec{q}*}_{m}(\vec{r})\op{O}(\vop{p})  \phi^{\vec{k}}_n(\vec{r})d\vec{r}\\
    =&\sum_{\vec{R}}\int_{\uc}
    u^{\vec{q}*}_{m}(\vec{r})e^{-i\vec{q}\cdot(\vec{r}+\vec{R})}
    \op{O}(\vop{p})
    \left[u^{\vec{k}}_n(\vec{r})e^{i\vec{k}\cdot(\vec{r}+\vec{R})} \right] d\vec{r}\\
    =&\sum_{\vec{R}}e^{-i(\vec{q}-\vec{k})\cdot\vec{R}}
    \int_{\uc}u^{\vec{k'}*}_{m}(\vec{r})e^{-i\vec{q}\cdot\vec{r}}\op{O} (\vop{p})
    \left[ u^{\vec{k}}_n(\vec{r})e^{i\vec{k}\cdot\vec{r}} \right]d\vec{r}\\
    =&N\delta_{\vec{q},\vec{k}}
    \int_\uc\phi^{\vec{q}*}_{m}(\vec{r})\op{O}(\vop{p})  \phi^{\vec{k}}_n(\vec{r})d\vec{r}\\
    =&N \delta_{\vec{q},\vec{k}}
    \bra{\phi^{\vec{q}}_{m}}\op{O}(\vop{p}) \ket{ \phi^{\vec{k}}_n}_\uc,
  \end{aligned}
\end{equation}
where we have used Eq.~\eqref{eq:ham_1d}.
For $\op{O}\equiv \op{1}$ in Eq.~\eqref{eq:ham_1f},
\begin{equation}
  \label{eq:struc_lat_9}
  \braket{\phi^{\vec{q}}_{m}}{\phi^{\vec{k}}_n}_\crys
  = N \delta_{\vec{q},\vec{k}} \braket{u^{\vec{q}}_m}{u^{\vec{k}}_n}_\uc
  \equiv N \delta_{\vec{q},\vec{k}} \delta_{mn},
\end{equation}
where the last equality represents the choice of orthonormality for the cell-periodic functions. Thus in our convention, the resolution of identity is
\begin{equation}
  \label{eq:struc_lat_9a}
  1 = N^{-1} \sum_{n\vec{k}}\ket{\phi_n^{\vec{k}}}\bra{\phi_n^{\vec{k}}}_\crys = \sum_{n\vec{k}} \ket{u_n^{\vec{k}}}\bra{u_n^{\vec{k}}}_\uc.
\end{equation}

Similarly, to calculate the matrix elements of the position operator $\vop{r}$ in the Bloch basis one would go  through similar steps as in Eq.~\eqref{eq:ham_1f}. However, the third equality would have an extra term involving the factor $\sum_{\vec{R}}\vec{R}e^{-i(\vec{q}-\vec{k})\cdot\vec{R}}$, which is difficult to evaluate and depends on the choice of the Bravais lattice. To circumvent this, we first consider the continuum limit where the super-cell volume goes to infinity, such that $\vec{k}$-sums turn into integrals, and Kronecker delta into delta functions,
\begin{align}
  \label{eq:struc_lat_10}
  N^{-1}\sum_{\vec{k}\in BZ} &\rightarrow \frac{\vuc}{(2\pi)^D}\int_{BZ}d\vec{k} \\
  N\delta_{\vec{q},\vec{k}} &\rightarrow\frac{(2\pi)^D}{\vuc}\delta(\vec{q}-\vec{k}).
\end{align}
Henceforth, we mostly use the discrete notation, but when $\vec{k}$-derivatives are involved the continuous case in Eq.~\eqref{eq:struc_lat_10} is implied.
In contrast to the momentum operator in Eq.~\eqref{eq:ham_1f}, the position operator couples Bloch states in $\vec{k}$-space \cite{Blount1962book},
\begin{equation}
  \label{eq:struc_lat_11}
  \begin{aligned}
   \bra{\phi_m^{\vec{q}}}\vop{r}\ket{\phi_n^{\vec{k}}}_\crys
   &=\int_\crys d\vec{r}\phi_m^{\vec{q}*}(\vec{r})\vec{r}\phi_n^{\vec{k}}(\vec{r})\\
   &=\int_\crys d\vec{r} \left\{
     i\nabla_{\vec{q}}\phi_m^{\vec{q}*}(\vec{r})
     -i\nabla_{\vec{q}}u_m^{\vec{q}*}(\vec{r})e^{-i\vec{q}\cdot\vec{r}}
   \right\}\phi_n^{\vec{k}}(\vec{r})\\
   &=i \nabla_{\vec{q}} \int_\crys d\vec{r}\phi_m^{\vec{q}*}(\vec{r})\phi_n^{\vec{k}}(\vec{r})
   -i \int_\crys d\vec{r}\nabla_{\vec{q}}u_m^{\vec{q}*}(\vec{r})e^{-i\vec{q}\cdot\vec{r}}
   \phi_n^{\vec{k}}(\vec{r})\\
   &=i N \delta_{m,n}\nabla_{\vec{q}}\delta_{\vec{q},\vec{k}}
   -i
    \sum_{\vec{R}} e^{-ij(\vec{q}-\vec{k})\cdot\vec{R}}
    \int_{\text{cell}}d\vec{r} 
    e^{-i(\vec{q}-\vec{k})\cdot\vec{r}}\nabla_{\vec{q}}u_m^{\vec{q}*}(\vec{r})u_n^{\vec{k}}(\vec{r})\\
   &= N(i\delta_{m,n}\nabla_{\vec{q}}
   + \vec{d}_{mn}^{\vec{k}} )\delta_{\vec{q},\vec{k}}
 \end{aligned}
\end{equation}
where we in the last step have used Eq.~\eqref{eq:ham_1d}, performed partial integration of the second term, and defined the generalized dipole coupling
\begin{equation}
  \label{eq:struc_lat_12}
  \begin{aligned}
    \vec{d}_{mn}^{\vec{k}}
    \equiv i\bra{u_m^{\vec{k}}}\nabla_{\vec{k}}\ket{u_n^{\vec{k}}}_\uc.
  \end{aligned}
\end{equation}
By taking the derivative of the TISE, it can be shown that  $\vec{d}_{mn}^{\vec{k}}=-i\vec{p}_{mn}^{\vec{k}} /(E_m^{\vec{k}}-E_n^{\vec{k}})$ for the nondiagonal elements. The diagonal elements are the Berry connections [see Eq.~\eqref{eq:adiab_9}],
\begin{equation}
  \label{eq:struc_lat_13}
  \vec{\berryc}_n^{\vec{k}} \equiv \vec{d}_{nn}^{\vec{k}}.
\end{equation}
Clearly, to be able to calculate $\vec{k}$-gradients in Eqs.~\eqref{eq:struc_lat_12} and \eqref{eq:struc_lat_13}, the cell-periodic functions $\ket{u_n^{\vec{k}}}$ should be smooth and BZ-periodic functions. Generally, $\vec{k}$-gradients are often encountered in the description of dynamics in solids, and it is thus desirable to ensure that $\ket{u_n^{\vec{k}}}$ fulfills the mentioned properties. This will be the topic of the subsequent subsections.

\subsection{Random structure gauge} \label{sec:ham_random}

As discussed above, the Bloch states in the TISE of Eq.~\eqref{eq:ham_1} are defined up to an arbitrary $\vec{k}$-dependent phase factor $e^{i\varphi_n^\vec{k}}$, i.e. there is a gauge degree of freedom. Since this gauge is related to the time-independent problem and the structure of the crystal,
we will refer to it as the structure gauge.
When Eq.~\eqref{eq:ham_1} is solved by some diagonalization procedure, the value of $\varphi_n^{\vec{k}}$ is usually random - we call this the \textit{random gauge}.

We consider first the non-degenerate case, and will discuss the degenerate case in Sec.~\ref{sec:ham_deg}.
The band energies have the BZ-periodicity $E_n^{\vec{k}}=E_n^{\vec{k}+\vec{G}}$, with $\vec{G}$ a reciprocal lattice vector, and the BZ can be regarded as a torus. 
With the gauge freedom, an ideal fixed gauge for our purposes is one in which the Bloch functions are BZ-periodic, as such a gauge would lead to BZ-periodic momentum and dipole matrix elements useful for numerical evaluations. The periodic gauge condition reads $\ket{\phi_n^{\vec{k}+\vec{G}}} = \ket{\phi_n^{\vec{k}}}$,
with $\vec{G}=\sum_{d=1}^{D}n_d\vec{b}_d$ an arbitrary reciprocal lattice vector, $D$ the dimension, $n_d$ integers, and $\vec{b}_d$ the primitive reciprocal lattice vectors. This condition is equivalent to
\begin{equation}
  \label{eq:ham_3}
  \ket{\phi_n^{\vec{k}+\vec{b}_d}} = \ket{\phi_n^{\vec{k}}}.
\end{equation}
We outline here a procedure to construct such a gauge, by first constructing a PT gauge, and afterwards constructing the TPT gauge with the Berry phases along the reciprocal vectors distributed evenly across the BZ.

\subsection{Parallel transport structure gauge} \label{sec:ham_pt}
Let us denote the Bloch states in the random gauge by $\ket{\rand{\phi}_n^{\vec{k}}}$ and the corresponding cell-periodic functions by $\ket{\rand{u}_n^{\vec{k}}}$.
In the PT gauge, similar as was discussed for Eq.~\eqref{eq:adiab_10} in Sec.~\ref{sec:berry}, the scalar Berry connections along the reduced coordinates [see Eq.~\eqref{eq:ham_1b}] are forced to vanish,
\begin{equation}
  \label{eq:ham_4}
  \pt{\berryc}_{n,\kappa_d}^{\vec{k}} \equiv i\bra{\pt{u}_n^{\vec{k}}}\partial_{\kappa_d}\ket{\pt{u}_n^{\vec{k}}}_\uc = 0,
\end{equation}
where the integral is carried over a unit cell. Note that to fix the gauge in Eq.~\eqref{eq:ham_4}, the cell-periodic functions $\ket{u_n^{\vec{k}}}$ are used over the Bloch states $\ket{\phi_n^{\vec{k}}}$ to define the scalar Berry connections - using the latter would have resulted in ambiguous integrals with fast oscillating integrands and dependence on the choice of spatial coordinate origin.
When fixing the gauge any condition can be used, it just turns out that using the cell-periodic functions are more convenient.
The full Berry connection vector can be constructed from the scalar Berry connections as
\begin{equation}
  \label{eq:ham_5}
  \vec{\berryc}_n^{\vec{k}}
  \equiv i\bra{u_n^{\vec{k}}}\nabla_{\vec{k}}\ket{u_n^{\vec{k}}}_\uc
  =\sum_{d,d'}^D \berryc_{n,\kappa_{d}}^{\vec{k}}g^{d,d'}\uv{b}_{d'},
\end{equation}
with $g^{d,d'}$ the inverse metric tensor.
In the discrete picture pertinent to numerical evaluations, the Berry connections in the PT gauge can be evaluated as \cite{Resta1994review}
\begin{equation}
  \label{eq:ham_4a}
  \pt{\berryc}_{n,\kappa_d}^{\vec{k}}=-\norm{\delta\vec{k}}^{-1}\Im\ln\braket{\pt{u}_n^{\vec{k}}}{\pt{u}_n^{\vec{k}+\delta\vec{k}}}_\uc,
\end{equation}
with $\delta\vec{k}$ a small displacement vector in reciprocal space.

\subsection{Periodic structure gauge} \label{sec:ham_tpt}
To construct the periodic gauge, note first that the periodic gauge condition for the Bloch states in Eq.~\eqref{eq:ham_3} translates into the condition
\begin{equation}
  \label{eq:ham_6}
  \ket{\tpt{u}_n^{\vec{k}+\vec{b}_d}} = e^{-i\vec{b}_d\cdot\vec{r}} \ket{\tpt{u}_n^{\vec{k}}}
\end{equation}
for the cell-periodic functions. The $\ket{\pt{u}_n^{\vec{k}}}$ constructed in the PT gauge are smooth inside the first BZ, but they generally do not satisfy Eq.~\eqref{eq:ham_6}.
For a closed path along $\vec{b}_d$ that wraps around the BZ, a Berry phase is accumulated, the so-called Zak's phase \cite{Zak1989PRL},
\begin{equation}
  \label{eq:ham_7}
  \varphi_{n,\kappa_d}^{B}=\oint_\text{BZ}\pt{\berryc}_{n,\kappa_d}d\kappa_d.
\end{equation}
It has been shown by Zak that for a one-dimensional solid with inversion symmetry, this phase can only take on the values 0 or $\pi$.

Consider now the discrete case, and assume that we have constructed the PT gauge along the $\uv{b}_d$ direction in discrete steps starting from one end of the BZ at $\vec{k}_0$, to the last point $\vec{k}_{N_d-1}$. From the PT gauge constraint in Eq.~\eqref{eq:ham_4}, the Berry connection along this path is identically zero. To close the loop of the integral in Eq.~\eqref{eq:ham_7}, we need to wrap around the BZ, and the Berry phase in the PT gauge is then calculated as
\begin{equation}
  \label{eq:ham_7a}
  \varphi^B_{n,\kappa_d} = - (\Delta \kappa_d)^{-1}\Im\ln \bra{\pt{u}_n^{\vec{k}_{N_d-1}}}e^{-i\vec{b}_d\cdot \vec{r}}\ket{\pt{u}_n^{\vec{k}_0}},
\end{equation}
where we have used Eq.~\eqref{eq:ham_6}, and $\Delta \kappa_d$ is the grid spacing of the reduced reciprocal coordinates defined in Eq.~\eqref{eq:ham_1b}.
Note that even though we defined the Berry phase in terms of the PT Berry connections, it is gauge-independent, in agreement with our discussions in Sec.~\ref{sec:berry}.
A periodic structure gauge is obtained by distributing this Berry phase evenly along the path onto the cell-periodic states in the PT gauge,
\begin{equation}
  \label{eq:ham_8}
  \ket{\tpt{u}_n^{\vec{k}}} \equiv e^{-i\varphi_{n,\kappa_d}^B\kappa_d/\norm{\vec{b}_d}}\ket{\pt{u}_n^{\vec{k}}}.
\end{equation}
For 2D or 3D systems ($D=2,3$), the above procedure is then repeated along all the dimensions $d$.
This gauge is denoted as the TPT gauge, and constitutes a periodic gauge with optimally smooth phase variation of the Bloch states \cite{Vanderbilt2018book}.

It is important to mention that a globally smooth periodic gauge is only possible for topologically trivial systems. For example, the Chern theorem states that the Berry phase [see Eq.~\eqref{eq:adiab_11}] calculated over a closed surface $S$ is quantized as a integer multiple of $2\pi$, which in the case of the Bloch problem reads, 
\begin{equation}
  \label{eq:ham_9}
  C_n = (2\pi)^{-1}\oiint_S \vec{\Omega}_n^{\vec{k}} \cdot d\vec{S},
\end{equation}
with $\vec{\Omega}_n^{\vec{k}}$ the Berry curvature [Eq.~\eqref{eq:adiab_12}], and $C_n$ is called the Chern number or the topological invariant. For a 2D system where the surface is the whole BZ, a nonzero Chern number (topologically nontrivial systems) presents a topological obstruction to the construction of a globally smooth periodic structure gauge \cite{Vanderbilt2018book}. %page93
For example, for topological systems, one can always construct the procedure for the TPT gauge along one dimension, but when one subsequently constructs the TPT gauge along the other dimension, the periodicity along the first dimension will break down somewhere in the BZ.
As we will discuss later in Sec.~\ref{sec:tdse} on the time-dependent problem, depending on the laser gauge, there are ways to circumvent the construction of such a global periodic gauge for HHG calculations.

\subsection{Wannier gauge} \label{sec:ham_wannier}
Up to this point, we have considered the Bloch states, which are entirely delocalized spatially. Often, however, it is useful to consider a spatially localized basis mimicking that of atomic or molecular orbitals. 
The construction of a periodic gauge opens up the possibility of constructing the Wannier basis \cite{Wannier1937PR, Wannier1962review}, which consists of states localized on the individual lattice sites.
The Wannier states are defined as the $\vec{k}$-space Fourier transforms of the Bloch states,
\begin{equation}
  \label{eq:ham_wannier_1}
  \ket{w_n^{\vec{R}}}=\mathcal{F}\left\{ \ket{\phi_n^{\vec{k}}}\right\},
  \quad
  \text{with} \quad
  \mathcal{F}\left\{ \cdot \right\} \equiv \frac{V_{\text{cell}}}{(2\pi)^D}\int_{\text{BZ}}e^{-i\vec{k}\cdot\vec{R}} \{\cdot\} d\vec{k},
\end{equation}
and $\vec{R}$ a Bravais lattice vector.
As mentioned, the Wannier functions are localized, in the sense that $w_n^{\vec{R}}(\vec{r})=w_n^{\vec{0}}(\vec{r}-\vec{R})$, and $|w_n^{\vec{R}}(\vec{r})|\rightarrow 0$ for $|\vec{r}-\vec{R}|\rightarrow \infty$. 

Clearly, to perform the integral over the BZ in Eq.~\eqref{eq:ham_wannier_1}, the employed Bloch states in the integrand should be smooth and periodic, i.e. they should be precalculated in a periodic gauge. Since such a gauge choice is not unique, the resulting Wannier states depend on the choice of the periodic gauge.
As the TPT structure gauge is the optimally-smooth periodic gauge, the Wannier states constructed using the TPT-gauge Bloch states are optimally localized (with minimum spatial spread), and are called the maximally-localized Wannier functions (MLWF) \cite{Kohn1959PR, Marzari2012review}.

Independent of the gauge, the Wannier states are orthonormal $\braket{w_n^{\vec{R}}}{w_{n'}^{\vec{R}'}}_\crys=\delta_{nn'}\delta_{\vec{R},\vec{R}'}$, and the Hilbert spaces spanned by the Bloch states and the Wannier states are identical.
When working with Wannier states as a basis, the terminology ``Wannier gauge''\cite{Wang2006PRB, Marzari2012review} is used, as the Wannier states are generally not energy eigenstates. Instead, the Hamiltonian matrix elements in the Wannier basis are the Fourier transforms of the band energies,
\begin{equation}
  \label{eq:ham_wannier_2}
  \bra{w_n^{\vec{0}}} \op{H}_0 \ket{w_n^{\vec{R}}}_\crys=\mathcal{F} \left\{ E_n^{\vec{k}} \right\}.
\end{equation}
The inverse transform of Eq.~\eqref{eq:ham_wannier_2}, $E_n^{\vec{k}}=\bra{\phi_n^{\vec{k}}}\op{H}_0\ket{\phi_n^{\vec{k}}}_\crys = \sum_{\vec{R}}e^{i\vec{k}\cdot\vec{R}}\bra{w_n^{\vec{0}}}\op{H}_0\ket{w_n^{\vec{R}}}_\crys$, shows that the Wannier states considered as the localized basis in tight-binding models exactly reproduces the band energies. The position matrix elements in the Wannier gauge are well defined, and are just the Fourier transforms of the Berry connections,
\begin{equation}
  \label{eq:ham_wannier_3}
  \bra{w_n^{\vec{0}}}\vop{r}\ket{w_n^{\vec{R}}}_\crys
  = \mathcal{F}\left\{ \vec{\berryc}_n^{\vec{k}} \right\}.
\end{equation}
The Wannier centers, $\bra{w_n^{\vec{0}}}\vop{r}\ket{w_n^{\vec{0}}}_\crys$, are invariant with respect to a gauge change of the Bloch states\footnote{However, as mentioned, the spread of the Wannier states are gauge-dependent.}, and in 1D are given by $a\varphi_n^B/(2\pi)$, with $\varphi^{B}$ Zak's phase and $a$ the lattice constant.

\subsection{Structure gauges for degenerate bands} \label{sec:ham_deg}

The gauge discussion in the previous subsections can be generalized to the degenerate case \cite{Vanderbilt2018book}, which we sketch here for completeness.
Let $\{E_n^{\vec{k}}\}_J$ be a set of $J$ bands that are isolated from all other bands, in the sense that they have no degeneracies with the other bands anywhere in the BZ, but can have degeneracies within themselves. The Hilbert subspace spanned by the corresponding $J$ states $\ket{u_n^{\vec{k}}}$ remains unchanged under an unitary transformation
\begin{equation}
  \label{eq:ham_deg_1}
  \ket{\pt{u}_n^{\vec{k}}} = \sum_{m=1}^{J}U_{mn}^{\vec{k}}\ket{u_m^{\vec{k}}},
\end{equation}
and the objective is to pick $U^{\vec{k}}$ such that the transformed states have the desired properties of our structure gauge.

We first discuss the construction of the PT-gauge states $\{\pt{u}_n^{\vec{k}}\}_J$ from a random gauge $\{\rand{u}_n^{\vec{k}}\}_J$, and consider the discrete-$\vec{k}$ case which is illustrative and useful for numerical applications.
The procedure can be considered a generalization of Sec.~\ref{sec:ham_pt}.
Starting at $\vec{k}_0$ with $\ket{\pt{u}_m^{\vec{k}_0}} \equiv \ket{\rand{u}_m^{\vec{k}_0}}$, the overlap matrix with a neighbouring point $\vec{k}_1=\vec{k}_0+\delta\vec{k}$ can be decomposed by a singular value decomposition
\begin{equation}
  \label{eq:ham_deg_2}
  M_{mn}^{\vec{k}_0\vec{k}_1} \equiv \braket{\pt{u}_m^{\vec{k}_0}}{\rand{u}_n^{\vec{k}_1}} = (V \Sigma W^\dagger)_{mn},
\end{equation}
with $\Sigma$ a diagonal matrix with non-negative values, and $V$, $W$ unitary matrices. The difference between $\Sigma$ and the identity matrix is a measure of the difference between the Hilbert space spanned by $\{u_n^{\vec{k}_0}\}_J$ and $\{u_n^{\vec{k}_1}\}_J$.
Define now a new set of states by transforming the set of states from the random gauge,
\begin{equation}
  \label{eq:ham_deg_3}
  \ket{\pt{u}_n^{\vec{k}_1}} = \sum_{m=1}^{J}(WV^\dagger)_{mn}\ket{\rand{u}_m^{\vec{k}_1}}.
\end{equation}
The overlap matrix can be shown using Eqs.~\eqref{eq:ham_deg_2} and \eqref{eq:ham_deg_3} to be
$\braket{\pt{u}_m^{\vec{k}_0}}{\pt{u}_n^{\vec{k}_1}} =(V\Sigma V^\dagger)_{mn}$ and is now Hermitian and positive definite, and is ``as close as possible`` to the identity matrix, such that the constructed set of states $\{\pt{u}_n^{\vec{k}_0}\}_J$ can be considered as ``optimally aligned'' to $\{\pt{u}_n^{\vec{k}_1}\}_J$. The PT gauge of a local region in the BZ can now be constructed by repeating the above procedure to all other nearby points in the BZ.

We proceed to discuss a procedure to construct a periodic gauge in the multiband case, which is a generalization of the nondegenerate TPT gauge discussed in Sec.~\ref{sec:ham_tpt}. First construct the PT gauge with the above procedure, starting from $\vec{k}_0$ at one end of the BZ along the reciprocal vector direction $\uv{b}_d$ to the other end at $\vec{k}_{N-1}$. Even though the Hamiltonian at $\vec{k}_0$ and $\vec{k}_N$ are identical, the set of states $\{\pt{\phi}_n^{\vec{k}_0}\}_J$ and $\{\pt{\phi}_n^{\vec{k}_N}\}_J$ generally are not. In analogy with the nondegenerate case, when wrapping around the BZ, we can define the unitary overlap matrix
\begin{equation}
  \label{eq:ham_deg_4}
  \mathcal{U}_{mn}=(\Delta\kappa_d)^{-1}\bra{\pt{u}_m^{\vec{k}_{N-1}}}e^{-i\vec{b}_d\cdot \vec{r}}\ket{\pt{u}_n^{\vec{k}_0}},
\end{equation}
from which we can define a Berry phase
\begin{equation}
  \label{eq:ham_deg_5}
  \varPhi^B_{\kappa_d}=-\Im\ln \det \mathcal{U}=\sum_{n=1}^{J}\varphi^B_{n,\kappa_d}.
\end{equation}
The $\varphi_{n,\kappa_d}^B$ are the argument of the eigenvalues of $\mathcal{U}$, and interpreted as the Berry phases for the individual bands.

Similar to the single-band case discussed in Sec.~\ref{sec:ham_tpt}, a periodic gauge can then be constructed from the PT gauge by dividing the Berry phases evenly along the path
\begin{equation}
  \label{eq:ham_deg_6}
  \ket{\tpt{u}_n^{\vec{k}}}=e^{-i\varphi_{n,\kappa_d}^B\kappa_d/\norm{\vec{b}_d}} \ket{\pt{u}_n^{\vec{k}}},
\end{equation}
which is the TPT gauge for the multiband case. It should be noted that the states in the multiband PT and TPT gauges, $\{\pt{\phi}_n^{\vec{k}}\}_J$ and $\{\tpt{\phi}_n^{\vec{k}}\}_J$, are generally not energy eigenstates.

Once the multi-band TPT gauge is constructed, the Wannier gauge in the multiband case can then be constructed from the TPT gauge as in the single-band case, i.e. using Eq.~\eqref{eq:ham_wannier_1}. The Wannier functions retain most of the important properties from the single-band case, such as localization in real space and lattice-periodicity \cite{Marzari2012review}.

\section{The time-dependent problem and laser-gauges} \label{sec:tdse}
In this section, we discuss the different avenues to tackle the time-dependent problem of a solid interacting with a strong laser field, by treating the crystal quantum mechanically and the external field classically. Depending on the chosen laser gauge and basis, the resulting EOMs will have their own advantages and drawbacks. The section starts with the introduction to laser gauge freedom, and proceeds in Secs.~\ref{sec:tdse_vg}-\ref{sec:tdse_houston} to obtain the relevant EOMs and equations for the microscopic current.
Section~\ref{sec:tdse_compare} compares the different time-dependent methods, in terms of both the numerical complexity and the interpretation of the physics, and Section~\ref{sec:tdse_hhg} describes the calculation of the HHG spectral and temporal profiles and provides an example.

The minimal coupling Hamiltonian for a nonrelativistic electron in a periodic potential interacting with an electromagnetic field reads \cite{Sakurai1994book, Bransden03book}
\begin{equation}
  \label{eq:tdse_1}
  \op{H}(t) = \frac{1}{2}[\op{\vec{p}}+\vec{A}(\vec{r}, t)]^2 - \Phi(\vec{r}, t) + V(\vec{r}), 
\end{equation}
with $\vec{A}(\vec{r},t)$ the vector potential, $\Phi(\vec{r}, t)$ the electric scalar potential, and the physical fields given by
\begin{align}
  \vec{F}(\vec{r}, t) &= -\nabla \Phi(\vec{r}, t) - \partial_t\vec{A}(\vec{r}, t) \label{eq:tdse_2} \\
  \vec{B}(\vec{r}, t) &= \nabla \times \vec{A}(\vec{r}, t).\label{eq:tdse_2b}
\end{align}
There is a gauge freedom in choosing $\vec{A}$ and $\Phi$, as the physical fields remain invariant under the transformations
\begin{subequations}
  \label{eq:tdse_3}
  \begin{align}
    \vec{A}(\vec{r},t) &\rightarrow \vec{A}(\vec{r}, t)+\nabla\Lambda(\vec{r}, t) \\
    \Phi(\vec{r},t) &\rightarrow \Phi(\vec{r}, t)-\partial_t\Lambda(\vec{r}, t)
  \end{align}
\end{subequations}
with $\Lambda(\vec{r}, t)$ a differentiable real function.
It is straightforward to show that under the gauge transform \eqref{eq:tdse_3}, the TDSE remains invariant if the wave function transforms as
\begin{align}
  \Psi(\vec{r}, t) & \rightarrow  \Psi'(\vec{r}, t) = e^{-i\Lambda(\vec{r},t)} \Psi(\vec{r}, t) \label{eq:tdse_5} \\
  \op{H}(t) & \rightarrow \op{H}'(t) = \frac{1}{2}[\op{\vec{p}}+\vec{A}(\vec{r}, t)+\nabla\Lambda(\vec{r},t)]^2 - \Phi(\vec{r}, t) + \partial_t \Lambda(\vec{r}, t) + V(\vec{r}), \label{eq:tdse_6}
\end{align}
where the Hamiltonian transform is explicitly written down in the second line, obtained by insertion of the gauge transform \eqref{eq:tdse_3} into the TDSE. Expectation values of physical observables such as the position $\bra{\Psi(t)}\vop{r}\ket{\Psi(t)}$ and kinetic momenta $\bra{\Psi(t)}\left[\vop{p}+\vec{A}(t)\right]\ket{\Psi(t)}$ are gauge invariant under the transform.

The microscopic current operator is proportional to the kinetic momentum
\begin{equation}
  \label{eq:tdse_7}
  \vop{j}(t) = -\vop{v}(t)=i\left[\vop{r}, \op{H}(t)\right] = -[\vop{p}+\vec{A}(\vec{r},t)],
\end{equation}
 where $[\cdot , \cdot]$ denotes a commutator.
The wavelengths of driving fields used for HHG are in the order of micrometers, while the unit cell dimension is sub-nanometer (nm), so here we apply the dipole approximation $\vec{F}(t)\equiv \vec{F}(\vec{r},t)$. In Sec.~\eqref{sec:macro} we will discuss situations beyond the dipole approximation. 
In addition, we ignore the the magnetic field (since the electrons are non-relativistic).

\subsection{Velocity gauge EOM in the Bloch basis} \label{sec:tdse_vg}
We start by considering the dynamics in the VG, in which one fixes the gauge by choosing $\Phi_\vg (\vec{r},t)=0$ and $\vec{A}_\vg (\vec{r},t) =-\int^t\vec{F}(t')dt'\equiv \vec{A}(t)$. As we will show, the VG can be advantangeous because it leads to EOMs that can be propagated separately for each $\vec{k}$, which means that the construction of a periodic structure gauge prior to time-propagation is unnecessary.
The VG Hamiltonian reads
\begin{equation}
  \label{eq:tdse_vg_1}
  \op{H}_\vg (t) = \frac{1}{2}[\op{\vec{p}}+\vec{A}(t)]^2 + V(\vec{r}).
\end{equation}
The $\vec{A}^2/2$ term can be transformed away by choosing $\Lambda(\vec{r},t) = -\frac{1}{2}\int^t\vec{A}(t')^2dt'$ in Eqs.~\eqref{eq:tdse_5} and \eqref{eq:tdse_6}. The resulting VG Hamiltonian form,
\begin{equation}
  \label{eq:tdse_vg_2}
  \op{H}_\vg '(t) = \op{T} + V(\vec{r}) + \op{\vec{p}}\cdot\vec{A}(t),
\end{equation}
is often used instead of $\op{H}_\vg$ in Eq.~\eqref{eq:tdse_vg_2}.

The VG TDSE $i\ket{\dot{\Psi}_\vg (t)} = \op{H}'_\vg(t)\ket{\Psi_\vg(t)}$ can be rewritten by first expanding the wave function
$\Psi_\vg(\vec{r}, t) = N^{-1} \sum_{m,\vec{k}\in BZ}b_m^{\vec{k}}(t)\phi_m^{\vec{k}}(\vec{r})$
and then projecting onto the Bloch states, resulting in the EOM for the coefficients \cite{Korbman2013NJP, Wu2015PRA}
\begin{equation}
  \label{eq:tdse_vg_3}
  i \dot{b}_m^{\vec{k}}(t) = E_m^{\vec{k}} b_m^{\vec{k}}(t) + \vec{A}(t) \cdot \sum_n \vec{p}_{mn}^{\vec{k}}b_n^{\vec{k}}(t),
\end{equation}
where we have used the property that the momentum matrix elements are diagonal in $\vec{k}$ [Eq.~\eqref{eq:ham_1f}], and defined $\vec{p}_{mn}^{\vec{k}}\equiv \bra{\phi_m^{\vec{k}}} \op{\vec{p}} \ket{\phi_n^{\vec{k}}}_\uc$.
We can define the density matrix operator $\op{g}(t)=\ket{\Psi_\vg(t)}\bra{\Psi_\vg(t)}$, such that the matrix elements $g_{mn}^{\vec{k}}\equiv b_m^{\vec{k}}b_n^{\vec{k}*}$ satisfy the EOM \cite{Yue2020PRA}
\begin{equation}
  \label{eq:tdse_vg_4}
  \begin{aligned}
    i\dot{g}_{mn}^{\vec{k}}(t)
    =&\left(E_m^{\vec{k}}-E_n^{\vec{k}}\right) g_{mn}^{\vec{k}}(t)
    +\vec{A}(t)\cdot \sum_l 
    \left[\vec{p}_{ml}^{\vec{k}} g_{ln}^{\vec{k}}(t) 
      - \vec{p}_{ln}^{\vec{k}} g_{ml}^{\vec{k}}(t)  \right].
  \end{aligned}
\end{equation}
These equations can also be derived directly from the Liouville-von Neumann equation $i\dot{\op{g}}(t)=[\op{H}(t),\op{g}(t)]$. 
Finally, the relevant observable for HHG, the microscopic current, is evaluated in the VG as
\begin{equation}
  \label{eq:tdse_vg_5}
  \begin{aligned}
    \vec{j}(t)
    & = \tr\left[\vop{j}_\vg (t)\op{g}(t)\right]
    = -\tr\left\{\left[\vop{p}+\vec{A}(t)\right]\op{g}(t)\right\} \\
    & = - N^{-1}\sum_{\vec{k}\in BZ}\sum_{mn}
    \left[ \vec{p}_{mn}^{\vec{k}} + \delta_{mn}\vec{A}(t) \right] g_{nm}^{\vec{k}}(t),
  \end{aligned}
\end{equation}
where we have used Eq.~\eqref{eq:tdse_7}.
We will discuss the advantages and drawbacks of time propagation in the VG  in Sec.~\ref{sec:tdse_compare}.

\subsection{Length gauge EOM in Bloch basis} \label{sec:tdse_lg}
A different way to fix the laser gauge in the dipole approximation is to set $\Phi_\lg(\vec{r},t)=-\vec{r}\cdot\vec{F}(t)$ and $\vec{A}_\lg=\vec{0}$, which according to Eq.~\eqref{eq:tdse_2} results in the desired physical field $\vec{F}(t)$. This gauge choice is denoted as the LG. The LG Hamiltonian is then according to Eq.~\eqref{eq:tdse_1}
\begin{equation}
  \label{eq:tdse_lg_1}
  \op{H}_\lg(t) = \op{T} + V(\vec{r}) + \vec{r}\cdot\vec{F}(t).
\end{equation}
This Hamiltonian can also be obtained from the VG Hamiltonian using the gauge transform $\Lambda_{VG \rightarrow LG}=-\vec{A}(t)\cdot\vec{r}$ in Eqs.~\eqref{eq:tdse_5} and \eqref{eq:tdse_6}.

We first rewrite the LG TDSE $i\ket{\dot{\Psi}_\lg (t)} = \op{H}_\lg(t)\ket{\Psi_\lg(t)}$ by expanding $\Psi_\lg(\vec{r}, t) = N^{-1}\sum_{m,\vec{k}\in BZ}a_m^{\vec{k}}(t)\phi_m^{\vec{k}}(\vec{r})$
and projecting onto the Bloch states, resulting in the EOM \cite{Souza2004PRB, Vampa2014PRL, Ernotte2018PRB}
\begin{equation}
  \label{eq:tdse_lg_2}
  i \dot{a}_m^{\vec{k}}(t) = E_m^{\vec{k}} a_m^{\vec{k}}(t)
  + \vec{F}(t) \cdot \sum_n \vec{d}_{mn}^{\vec{k}} a_n^{\vec{k}}(t)
  + i \vec{F}(t) \cdot \nabla_{\vec{k}}a_m^{\vec{k}}(t).
\end{equation}
In contrast to the VG in Eq.~\eqref{eq:tdse_vg_3}, the LG equations couple different crystal momenta to each other due to the last term involving the $\vec{k}$-gradient.
Defining the LG density matrix $\op{\rho}(t)=\ket{\Psi_\lg(t)}\bra{\Psi_\lg(t)}$ with matrix elements $\rho_{mn}^{\vec{k}}=a_m^{\vec{k}}a_n^{\vec{k}*}$, the density matrix EOM reads
\begin{equation}
  \label{eq:tdse_lg_3}
  \begin{aligned}
    i\dot{\rho}_{mn}^{\vec{k}}(t)
    =&\left(E_m^{\vec{k}}-E_n^{\vec{k}}\right) \rho_{mn}^{\vec{k}}(t)
    +\vec{F}(t)\cdot \sum_l 
    \left[\vec{d}_{ml}^{\vec{k}} \rho_{ln}^{\vec{k}}(t) 
      - \vec{d}_{ln}^{\vec{k}} \rho_{ml}^{\vec{k}}(t)  \right]
    +i\vec{F}(t)\cdot\nabla_{\vec{k}}\rho_{mn}^{\vec{k}}(t).
  \end{aligned}
\end{equation}
As in the VG, these LG equations can also be derived from the Liouville-von Neumann equation $i\dot{\op{\rho}}(t)=[\op{H}_\lg(t),\op{\rho}(t)]$ in the Bloch basis.
In a many-body framework using second quantization, a similar equation for the reduced density matrix $\rho_{mn}^{\vec{k}}(t)\equiv\bra{\Psi(t)}a_{n\vec{k}}^\dagger a_{m\vec{k}}\ket{\Psi(t)}$ can be derived (with $a_{n\vec{k}}^\dagger$ creation and $a_{m\vec{k}}$  annihilation operators), often denoted as the semiconductor Bloch equations (SBEs) \cite{Sipe2000PRB, Haug2004book, Virk2007PRB, Golde2008PRB, Kira2012book, Schubert2014NPhoton, Ventura2017PRB}. Due to many-body couplings such as electron-electron and electron-phonon scattering, the SBEs will dephase, which at our level of theory is treated by adding a phenomenological dephasing term $(1-\delta_{mn})\rho_{mn}^{\vec{k}}/T_2$ on the right-hand side of Eq.~\eqref{eq:tdse_lg_3}. A shorter dephasing time $T_2$ will result in less noisy HHG spectra, and $T_2$ is often chosen such that there is reasonable agreement between experiment and theory \cite{Vampa2014PRL, Vampa2015Nature, Lu2019NPhoton}. The phenomenogical dephasing, while being computationally convenient, can only give qualitative results, and more accurate treatments of dephasing is an active area of research \cite{Floss2018PRA, Floss2019PRB, Kilen2020PRL, Du2018PRAb}.

For degenerate subspaces given in the periodic gauge [see Sec.~\ref{sec:ham_deg}], the rotated states of Eq.~\eqref{eq:ham_deg_1} are not energy eigenstates, and the EOM given in Eq.~\eqref{eq:tdse_lg_3} cannot be applied. In this case, the Liouville-von Neumann expression should be used to reduce the correct propagation equations \cite{Virk2007PRB, Thong2021PRB}. The correct propagation equations in the Wannier structure gauge (see Sec.~\ref{sec:ham_wannier}) can also be derived using the same strategy \cite{Ventura2017PRB, Silva2019PRB}.

In the LG, the current operator is $\vop{j}_\lg(t)=-\vop{p}$ [Eq.~\eqref{eq:tdse_7}], and the microscopic current can be evaluated as
\begin{equation}
  \label{eq:tdse_lg_4}
  \begin{aligned}
    \vec{j}(t)
    & = \tr\left[\vop{j}_\lg(t)\op{\rho}(t)\right]
    & = - N^{-1}\sum_{\vec{k}\in BZ}\sum_{mn}\vec{p}_{mn}^{\vec{k}}\rho_{nm}^{\vec{k}}(t).
  \end{aligned}
\end{equation}
While Eq.~\eqref{eq:tdse_lg_4} is useful to obtain the total current, more physical insight can be gained by a decomposition of the current into several terms with different physical meanings. While the decomposition into intraband and interband currents have been discussed extensively in the literature, we present a decomposition here that contains four terms each with its own physical interpretation \cite{Aversa1995PRB, Wilhelm2021PRB}.
We first split the position operator into an intraband and an interband component, $\vop{r}=\vop{r}^\tra+\vop{r}^\ter$, with the matrix elements in the Bloch basis [see Eqs.~\eqref{eq:struc_lat_11}-\eqref{eq:struc_lat_13}] 
\begin{subequations}
  \label{eq:tdse_lg_5}
  \begin{align}   
    \bra{\phi_m^{\vec{k}}}\op{\vec{r}}^\tra\ket{\phi_n^{\vec{q}}}_\crys
    & = N \delta_{mn}(\vec{\berryc}_m^{\vec{k}} + i\nabla_{\vec{k}})\delta_{\vec{k}\vec{q}} \\
    \bra{\phi_m^{\vec{k}}}\op{\vec{r}}^\ter\ket{\phi_n^{\vec{q}}}_\crys
    & = N (1 - \delta_{mn})\delta_{\vec{k}\vec{q}}\vec{d}_{mn}^{\vec{k}}.
  \end{align}
\end{subequations}
The current can now be written 
\begin{equation}
  \label{eq:tdse_lg_6}
  \begin{aligned}
    & \vec{j}(t) = \text{Tr}\left\{ \vop{j}_\lg (t) \op{\rho}(t) \right\}
    = i \text{Tr}\left\{ [\vop{r},\op{H}_\lg (t)] \op{\rho}(t) \right\} \\
    = & i \text{Tr}\left( \left\{
        [\vop{r}^\tra,\op{H}_0]
        + [\vop{r}^\tra, \vec{F}(t) \cdot \vop{r}^\tra]
        + [\vop{r}^\tra, \vec{F}(t) \cdot \vop{r}^\ter]
        + [\vop{r}^\ter,\op{H}_\lg (t)] \right\}
      \op{\rho} \right) \\
    \equiv & \vec{j}^\tra(t) + \vec{j}^\anom(t) + \vec{j}^\mix(t) + \vec{j}^\ter(t)
  \end{aligned}
\end{equation}
where we have inserted the LG Hamiltonian \eqref{eq:tdse_lg_1}.
After some tedious, but straight-forward derivations using Eqs.~\eqref{eq:ham_1} and \eqref{eq:tdse_lg_5}, the four current components in Eq.~\eqref{eq:tdse_lg_6} can be written as \cite{Aversa1995PRB},
\begin{subequations}
  \label{eq:tdse_lg_7}
  \begin{align}
    \vec{j}^\tra(t) &= - N^{-1}\sum_{m\vec{k}} \nabla_{\vec{k}}E_m^{\vec{k}} \rho_{mm}^{\vec{k}}(t) \label{eq:tdse_lg_7a} \\
    \vec{j}^\ter(t) &= - N^{-1}\partial_t \sum_{m\ne n, \vec{k}} \vec{d}_{mn}^{\vec{k}}
                    \rho_{nm}^{\vec{k}}(t) \label{eq:tdse_lg_7b} \\
    \vec{j}^\anom(t) &= - N^{-1} \sum_{m\vec{k}} \left[
                       \vec{F}(t) \times
                       \vec{\Omega}_m^{\vec{k}}
                     \right] \rho_{mm}^{\vec{k}}(t) \label{eq:tdse_lg_7c} \\
    \vec{j}^\mix(t) &= - N^{-1} \sum_\mu F_\mu(t) \sum_{m\ne n, \vec{k}}\left [
                      \left( \nabla_{\vec{k}} d_{\mu,mn}^{\vec{k}} \right)
                      - i ( \vec{\berryc}_{m}^{\vec{k}} - \vec{\berryc}_{n}^{\vec{k}}) d_{\mu,mn}^{\vec{k}}
                      \right] \rho_{nm}^{\vec{k}}(t). \label{eq:tdse_lg_7d},
  \end{align}
\end{subequations}
where $F_\mu(t)$ and $d_{\mu,mn}^{\vec{k}}$ are respectively the $\mu$th component of the field and dipole matrix elements. In Eq.~\eqref{eq:tdse_lg_7a}, the intraband current $\vec{j}^\tra(t)$ is due to carrier transport within individual bands as is reflected in its dependence on the carrier group velocities $\nabla_{\vec{k}}E_m^{\vec{k}}$. The interband current $\vec{j}^\ter(t)$ in Eq.~\eqref{eq:tdse_lg_7b} originates from the coupling between the bands and is seen to be just the time-derivative of the polarization. The anomalous current $\vec{j}^\anom(t)$ depends on the Berry curvature $\vec{\Omega}_m^{\vec{k}}=\nabla_{\vec{k}}\times \vec{\berryc}_m^{\vec{k}}$ [see Eq.~\eqref{eq:adiab_12}], and is perpendicular to the electric field $\vec{F}(t)$. The mixture current $\vec{j}^\mix(t)$ in Eq.~\eqref{eq:tdse_lg_7b}, as seen from the definition in Eq.~\eqref{eq:tdse_lg_6}, depends on the coupling between the interband and intraband position operators. The structure-gauge-invariant expression in the square parenthesis is also referred to as the generalized derivative \cite{Aversa1995PRB}, defined as $(O_{mn}^{\vec{k}})_{;\vec{k}}\equiv \nabla_{\vec{k}}O_{mn}^{\vec{k}} - i (\vec{\berryc}_m^{\vec{k}}-\vec{\berryc}_n^{\vec{k}})O_{mn}^{\vec{k}}$.
We note here that even though the $\vec{j}^\anom(t)$ contribution to the current originated in the intraband component of the position operator in Eq.~\eqref{eq:tdse_lg_6}, it and the mixture term $\vec{j}^\mix(t)$ depend explicitly on coherences between different bands and as such are inter-band in nature. 

More generally, it is worth mentioning that in the literature,  
the decomposition in terms of the intraband [Eq.~\eqref{eq:tdse_lg_7a}] and interband currents [Eq.~\eqref{eq:tdse_lg_7b}] has been discussed extensively for HHG in solids \cite{Schubert2014NPhoton, Vampa2014PRL, Vampa2015Nature, Luu2015Nature, Wang2017NCommun, Kaneshima2018PRL, Klemke2019NCommun, Yoshikawa2019NCommun}. The effect from the anomalous velocity and the Berry curvature has been investigated mostly separately \cite{Liu2017NPhys, Banks2017PRX, Luu2018NCommun}. The implication of the mixture terms in Eq.~\eqref{eq:tdse_lg_7d} has largely been unexplored \cite{Wilhelm2021PRB}. Often, the non-intraband current (containing the interband, anomalous and mixture currents) is used interchangeably with the interband current, either due to semantics, or because the anomalous and mixture currents often are negligible. Sometimes, the anomalous current is also considered as a part of the intraband current \cite{Silva2019NPhoton, Chacon2020PRB}, since it depends on the carrier density in a given band [see Eq.~\eqref{eq:tdse_lg_7c}] and can be considered a band-specific current\footnote{Remember however that the Berry curvature is a geometric property stemming from the residual coupling between the bands [see Eq.~\eqref{eq:adiab_13}].}.
The interplay between the four current contributions investigated under a combined framework could be a potential avenue of future research.

\subsection{EOM in the adiabatic Houston basis} \label{sec:tdse_houston}

We next discuss a popular propagation method that employs the Houston states, which, like the LG EOMs discussed in the previous subsection, can naturally incorporate a phenomenological description of dephasing. We first reformulate the time-dependent problem into one that draws parallel to the general adiabatic theory discussed in Sec.~\ref{sec:berry}. 
First realize that we can define a transformed time-independent Hamiltonian, $\op{H}_0^{\vec{k}}\equiv e^{-i\vec{k}\cdot\vec{r}}\op{H}_0 e^{i\vec{k}\cdot\vec{r}}$, with the cell-periodic functions as eigenstates and the same band energies as the Bloch states
\begin{equation}
  \label{eq:tdse_houston_1}
  \op{H}_0^{\vec{k}}u_n^{\vec{k}}(\vec{r}) = E_n^{\vec{k}}u_n^{\vec{k}}(\vec{r}).
\end{equation}
For the time-dependent VG Hamiltonian in Eq.~\eqref{eq:tdse_vg_1}, the transformed Hamiltonian can easily be shown to satisfy
\begin{equation}
  \label{eq:tdse_houston_2}
  \op{H}^{\vec{K}}(t) \equiv e^{-i\vec{K}\cdot\vec{r}}\op{H}_\vg(t) e^{i\vec{K}\cdot\vec{r}}  = \op{H}^{\vec{K}+\vec{A}(t)}_0.
\end{equation}
Using Eqs.~\eqref{eq:tdse_houston_1} and \eqref{eq:tdse_houston_2}, we see that a set of adiabatic states [Eq.~\eqref{eq:adiab_1}] of $\op{H}^{\vec{K}}(t)$ simply consists of the cell-periodic functions with shifted crystal momenta, $u_n^{\vec{K}+\vec{A}(t)}(\vec{r})$, with corresponding adiabatic eigenenergies $E_n^{\vec{K}}(t)=E_n^{\vec{K}+\vec{A}(t)}$.
The solution to the adiabatic problem of the original VG Hamiltonian can now be written down,
\begin{equation}
  \label{eq:tdse_houston_3}
  \begin{aligned}
    \op{H}^{\vec{K}}(t) u_n^{\vec{K}+\vec{A}(t)}(\vec{r})
    =& E^{\vec{K}+\vec{A}(t)} u_n^{\vec{K}+\vec{A}(t)}(\vec{r}) \\
    \Leftrightarrow
    \op{H}_\vg(t) h_n^{\vec{K}}(\vec{r},t)
    =& E^{\vec{K}+\vec{A}(t)} h_n^{\vec{K}}(\vec{r},t)
  \end{aligned}
\end{equation}
with
\begin{equation}
  \label{eq:tdse_houston_3b}
  h_n^{\vec{K}}(\vec{r},t)
  \equiv e^{i\vec{k}\cdot\vec{r}} u_n^{\vec{K}+\vec{A}(t)}(\vec{r})
  = e^{-i\vec{A}\cdot\vec{r}} \phi_n^{\vec{K}+\vec{A}(t)}(\vec{r}).
\end{equation}
This set of adiabatic states $h_n^{\vec{K}}(\vec{r},t)$ are often referred to as Houston states in the literature \cite{Houston1940PR, Krieger1986PRB}, and we have used capital letter $\vec{K}$ for the crystal momenta to highlight that the Houston states are the accelerated Bloch states.
All the properties of the adiabatic states discussed in Sec.~\ref{sec:berry} now automatically follow.
If the structure gauge for the Bloch states is fixed, then the Houston states are completely well-determined and given by Eq.~\eqref{eq:tdse_houston_3b}. 
However, we emphasize that the Houston states only represent one possible set of adiabatic states - at each instant of time $t$, a Houston state multiplied by an arbitrary $\vec{k}$-dependent phase factor is another adiabatic state. To rephrase, an adiabatic state of $\op{H}_\vg(t)$, $\ket{n(t)}$, is always related to a Houston state by an arbitrary phase factor
\begin{equation}
  \label{eq:tdse_houston_3c}
  \braket{\vec{r}}{n(t)}=e^{i\varphi_n^{\vec{k}}}h_n^{\vec{K}}(\vec{r},t).
\end{equation}
Since the adiabatic states are defined as the instantaneous eigenstates of the Hamiltonian involving the laser-matter interaction, they can also be considered as the laser-field-dressed states.

Expanding the wave function in the Houston basis, $\Psi_\vg(\vec{r}, t) = N^{-1}\sum_{m,\vec{k}}c_m^{\vec{K}}(t)h_m^{\vec{K}}(\vec{r},t)$, the VG TDSE reads
\begin{equation}
  \label{eq:tdse_houston_4}
  i \dot{c}_m^{\vec{K}}(t) = E_m^{\vec{K}+\vec{A}(t)} c_m^{\vec{K}}(t)
  + \vec{F}(t) \cdot \sum_n \vec{d}_{mn}^{\vec{K}+\vec{A}(t)} c_n^{\vec{K}}(t),
\end{equation}
where we have calculated the nonadiabatic couplings [see Eq.~\eqref{eq:adiab_4}]
 \begin{equation}
   \label{eq:tdse_houston_4b}
   i\braket{h_m^{\vec{Q}}(t)}{\dot{h}_n^{\vec{K}}(t)}_\crys
   =Ni\delta_{\vec{Q},\vec{K}}\braket{u_m^{\vec{Q}+\vec{A}(t)}}{\dot{u}_n^{\vec{K}+\vec{A}(t)}}_\uc
   = -N\delta_{\vec{Q},\vec{K}}\vec{F}(t) \cdot \vec{d}_{mn}^{\vec{K}+\vec{A}(t)}.
 \end{equation}
The corresponding density matrix elements in the Houston basis, $\bar{\rho}_{mn}^{\vec{K}}=c_m^{\vec{K}}c_n^{\vec{K}*}$, evolve as
\begin{equation}
  \label{eq:tdse_houston_5}
  \begin{aligned}
    i\dot{\bar{\rho}}_{mn}^{\vec{K}}(t)
    =&\left(E_m^{\vec{K}+\vec{A}(t)}-E_n^{\vec{K}+\vec{A}(t)}\right) \bar{\rho}_{mn}^{\vec{K}}(t)
    +\vec{F}(t)\cdot \sum_l 
    \left[\vec{d}_{ml}^{\vec{K}+\vec{A}(t)} \bar{\rho}_{ln}^{\vec{K}}(t) 
      - \vec{d}_{ln}^{\vec{K}+\vec{A}(t)} \bar{\rho}_{ml}^{\vec{K}}(t)  \right].
  \end{aligned}
\end{equation}
Similarly as in the previous subsection, for a degenerate subspace expressed in a periodic gauge, the Liouville-von Neumann equation need to be used to reduce the relevant EOMs.
Note that in the literature, due to the appearance of the dipole operator in the EOMs of Eq.~\eqref{eq:tdse_houston_5}, it has also been termed as the LG SBEs in the moving frame \cite{Vampa2014PRL,Floss2018PRA, Li2019PRA, Yue2020PRA}. Indeed, the EOMs in the VG adiabatic basis [Eqs.~\eqref{eq:tdse_lg_2} and \eqref{eq:tdse_lg_3}] can be obtained from the LG EOMs [Eqs.~\eqref{eq:tdse_houston_4} and \eqref{eq:tdse_houston_5}] by the frame change $\vec{k}=\vec{K}+\vec{A}(t)$. 
As in the LG case of Sec.~\ref{sec:tdse_lg}, dephasing can be introduced by a phenomenological term $(1-\delta_{mn})\bar{\rho}_{mn}^{\vec{K}}/T_2$ on the right-hand side of Eq.~\eqref{eq:tdse_houston_5}. We note that one can {\it not} add this dephasing term directly to EOM in the Bloch-basis VG [Eq.~\eqref{eq:tdse_vg_4}] due to the severe mixing of the field-free states in the presence of the laser field.

The microscopic current is evaluated as
\begin{equation}
  \label{eq:tdse_houston_6}
  \begin{aligned}
    \vec{j}(t)
    = \text{Tr}\left[ \vop{j}_\vg(t) \op{g}(t)\right]
    = -\tr\left\{\left[\vop{p}+\vec{A}(t)\right]\op{g}(t)\right\}
    = - N^{-1} \sum_{mn\vec{K}} \vec{p}_{mn}^{\vec{K}+\vec{A}(t)}  \bar{\rho}_{nm}^{\vec{K}}(t),
  \end{aligned}
\end{equation}
where we have used the identity
$\bra{h_m^{\vec{K}}(t)} \left[\vop{p}+\vec{A}(t)\right] \ket{h_n^{\vec{Q}}(t)}_\crys=\delta_{\vec{K}\vec{Q}}\vec{p}_{mn}^{\vec{K}+\vec{A}(t)}$. The current can be split into an intraband and an non-intraband contribution, by splitting the diagonal and off-diagonal terms of of the momentum matrix elements in Eq.~\eqref{eq:tdse_houston_6}, i.e. $\vec{j}(t) = \vec{j}_\tra(t) + \vec{j}_\ntra(t)$
\begin{subequations}
  \label{eq:tdse_houston_7}
  \begin{align}
    \vec{j}_\tra(t) = & - N^{-1} \sum_{n\vec{K}} \nabla_{\vec{K}}E_{n}^{\vec{K}+\vec{A}(t)} \bar{\rho}_{nn}^{\vec{K}} \label{eq:tdse_houston_7a} \\
    \vec{j}_\ntra(t) = & - N^{-1} \sum_{m\ne n,\vec{K}} \vec{p}_{mn}^{\vec{K}+\vec{A}(t)} \bar{\rho}_{nm}^{\vec{K}}, \label{eq:tdse_houston_7b}
  \end{align}
\end{subequations}
where the non-intraband current contains the interband, anomalous and mixture contributions to the current [compare to Eqs.~\eqref{eq:tdse_lg_6} and \eqref{eq:tdse_lg_7}].

\subsection{Gauge comparisons} \label{sec:tdse_compare}
\begin{figure}
  \centering
  \includegraphics[width=1.0\textwidth, clip, trim=0 0cm 0 0cm]{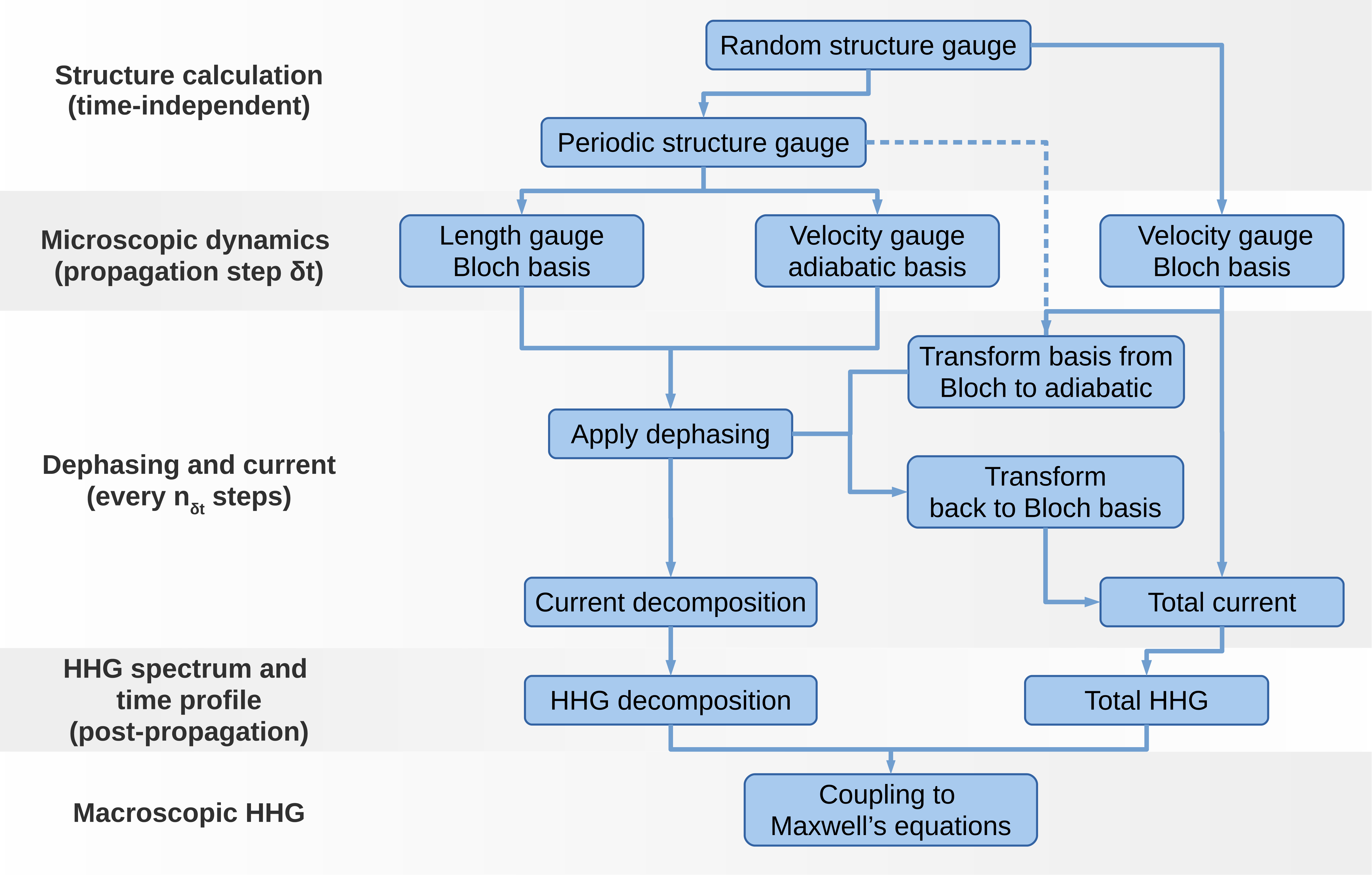}
  \caption{Flowchart sketching the computational steps for HHG in solids, consisting of a structure calculation, the propagation of the microscopic dynamics, the inclusion of dephasing and current, the spectrum calculation, and the macroscopic propagation.}
  \label{fig:tdse_compare_1}
\end{figure}

 The three different time-dependent propagation methods presented in Secs.~\ref{sec:tdse_vg}-\ref{sec:tdse_houston} each have their own advantages and drawbacks. Here we provide a discussion on this topic. The calculation procedures consist, in broad stokes, of five steps that are summarized in the flowchart of Fig.~\ref{fig:tdse_compare_1}.
These steps are: (1) Perform a structure calculation by some diagonalization procedure to obtain the band structure and coupling matrix elements. As discussed in Sec.~\ref{sec:ham_random}, the output will in general be in a random structure gauge. If required, we can then construct a periodic structure gauge using the procedures outlined in Secs.~\ref{sec:ham_pt}-\ref{sec:ham_deg}. (2) Next, we propagate the time-dependent EOMs, using one of the laser gauges discussed in Sec.~\ref{sec:tdse_vg}-\ref{sec:tdse_houston}. During the time propagation, every $n_{\delta t}$th steps, we can apply the dephasing and calculate the time-dependent currents. (4) After the time-propagation, with full knowledge of the time-dependent current, we can calculate the HHG spectrum and obtain the time-frequency information of the harmonics [see Sec.~\ref{sec:tdse_hhg}]. (5) Finally, in bulk crystals, the microscopic dynamics can be coupled with the Maxwell's equations to account for the macroscopic propagation of the laser and harmonics through the bulk, a topic which we will revisit in Sec.~\ref{sec:macro}.

Going through the flowchart in more detail, we start by considering a VG description of the dynamics (see right-hand side of flowchart). 
As discussed in Sec.~\ref{sec:tdse_vg}, the VG EOMs represented in the Bloch basis [see Eq.~\eqref{eq:tdse_vg_4}] are diagonal in the crystal momenta $\vec{k}$, meaning that each $\vec{k}$ can be propagated independently. Since no fixed phase relationship between neighbouring $\ket{\phi_n^{\vec{k}}}$ are required, there is no need to construct the periodic structure gauge (Sec.~\ref{sec:ham_tpt}) prior to the time-propagation, and using momentum matrix elements $\vec{p}_{mn}^{\vec{k}}$ obtained within a random structure gauge (see Sec~\ref{sec:ham_random}) is sufficient to calculate the total current in Eq.~\eqref{eq:tdse_vg_5} (see Fig.~\ref{fig:tdse_compare_1}). If desired, it is possible to include the phenomenological dephasing effect also into the VG Bloch basis calculation. 
This can be achieved during the time-propagation by first transforming into a VG adiabatic basis at desired times (separated by intervals $\Delta t=n_{\delta t}\delta_t$), applying the dephasing as
\begin{equation}
  \label{eq:tdse_compare_2}
  \bar{\rho}_{mn}^{\vec{k}} \rightarrow \bar{\rho}^{\vec{k}}_{mn}e^{-\Delta t/T_2}, \qquad m\ne n,
\end{equation}
and then transforming back to the Bloch basis. 
We here present two paths for such a basis transform. One method is to directly transform to the Houston basis \cite{Houston1940PR, Krieger1986PRB}
\begin{equation}
  \label{eq:tdse_compare_3}
  \begin{aligned}
    g_{mn}^{\vec{k}}(t) =& \sum_{l k} Q_{ml}^{\vec{k}}(t) \bar{\rho}_{lk}^{\vec{k}} Q_{nk}^{\vec{k}*}(t),
  \end{aligned}  
\end{equation}
with $Q_{mn}^{\vec{k}}(t)\equiv \braket{u_m^{\vec{k}}}{u_n^{\vec{k}+\vec{A}(t)}}_\uc $, and where we have used the resolution of identity and $\braket{\phi_m^{\vec{q}}}{h_n^{\vec{k}}(t)}_\crys=\delta_{\vec{q}\vec{k}} Q_{mn}^{\vec{k}}(t)$. This approach was used e.g. in Refs.~\cite{Foldi2017PRB, Ernotte2018PRB, Yue2020PRA}. Evaluation of Eq.~\eqref{eq:tdse_compare_3} presents several potential difficulties: it requires the direct knowledge of the cell-periodic functions $\ket{u_n^{\vec{k}}}$; the computational complexity is high \cite{Yue2020PRA}; and the periodic structure gauge is required since $\ket{u_n^{\vec{k}}}$ needs to be splined at $\vec{k}+\vec{A}(t)$\footnote{Unless an analytical form of $\ket{u_n^{\vec{k}}}$ is known, which is unlikely for real materials beyond simple model systems.} (dashed arrow in Fig.~\ref{fig:tdse_compare_1}). A different method to transform to an adiabatic basis is by using the definition in Eq.~\eqref{eq:adiab_1}, i.e. by diagonalizing the instantaneous Hamiltonian in the Bloch basis. As discussed in Eq.~\eqref{eq:tdse_houston_3c}, a resulting adiabatic state will differ from a Houston state by an arbitrary phase factor, which will not affect the inclusion of the phenomenological dephasing. While instantaneous diagonalization has similar computational complexity as Eq.~\eqref{eq:tdse_compare_3}, a random structure gauge is sufficient since each $\vec{k}$ can be treated independently. Note that such a forward and backward transform is not required at each propagation time step $\delta t$, but rather at time step $\Delta t=n_{\delta t}\delta_t$, with $n_{\delta t}\gg 1$, reducing the computational complexity. In the VG Bloch basis, compared to the other two discussed methods, more bands are often required to achieve convergence \cite{Aversa1995PRB, Virk2007PRB, Taghizadeh2017PRB, Foldi2017PRB, Yue2020PRA}, but at the same time degenerate bands can naturally be treated. In a future publication we will go into more details of treating HHG in the VG Bloch basis and decomposition of the current.

On the left-hand side of the flowchart, the LG approach involves the coupling of different crystal momenta by the $\nabla_{\vec{k}}$ term as discussed in Sec.~\ref{sec:tdse_lg}, due to the non-diagonal nature of the position operator $\vop{r}$ in the Bloch basis. 
Hence, prior to the time propagation, the construction of a periodic structure gauge is required. The phenomenological dephasing term can be naturally included during propagation, and the total current can be decomposed into the four terms given in Eq.~\eqref{eq:tdse_lg_6}. For the VG EOMs in the Houston basis described in Sec.~\ref{sec:tdse_lg}, a periodic structure gauge is also required since the dipole couplings need to be splined at crystal momenta $\vec{K}+\vec{A}(t)$. The calculation of the numerical gradient or the splining generally requires a much finer $\vec{k}$-space sampling than the VG Bloch scheme to obtain the same level of convergence, which severely increases the computational effort \cite{Virk2007PRB, Yue2020PRA}.
Similar to the LG case, the current can actually be decomposed into intraband and non-intraband terms, as presented in Eq.~\eqref{eq:tdse_houston_7}.
Compared to the VG Bloch basis case, the LG EOM in the Bloch basis and VG EOMs in the Houston basis often require a smaller number of bands for convergence.
Note that we have mostly discussed the construction of a globally smooth structure gauge, with the advantage that the gauge must only be constructed once, prior to the time-propagation. In the literature, a method exist \cite{Virk2007PRB} that aims to construct a locally smooth structure gauge during the time propagation of the LG SBEs, which has recently been applied to HHG \cite{Thong2021PRB}. For electric fields with changing polarization direction, the gauge construction in all three Cartesian directions must be applied at each time step, potentially severely increasing the computational complexity.
The key advantages of our three discussed propagation schemes are briefly summarized in Table.~\ref{tab:tdse_compare_1}.

\begin{table}[h]
  \small
  \centering
  \caption{Relative advantages of the VG Bloch basis propagation scheme compared to the LG Bloch basis or Houston basis propagation schemes.}
  \label{tab:tdse_compare_1}
  \begin{tabular}{m{4.cm} | m{4.cm} }
    \toprule
    \textbf{LG Bloch or Houston scheme}   
    & \textbf{VG Bloch scheme} \\ \midrule 
    Small number of bands for convergence.
    & No periodic structure gauge requirement.  \\ \hline
    Easier inclusion of dephasing.
    & Less number of $\vec{k}$-points for convergence. \\ \hline
    Easier current decomposition.
    & Easier treatment of degenerate bands. \\
      \bottomrule
  \end{tabular}
\end{table}

\subsection{HHG calculation and time profiles} \label{sec:tdse_hhg}

The HHG spectral yield is proportional to the spectral intensity of the current, and is calculated as Larmor's formula \cite{Jackson1999book}
\begin{equation}
  \label{eq:tdse_hhg_1}
  S(\omega) \propto \omega^2\norm{\vec{j}(\omega)}^2,
\end{equation}
with $\vec{j}(\omega)$ the Fourier transform of the time-dependent current,
\begin{equation}
  \label{eq:tdse_hhg_2}
  \vec{j}(\omega) = (2\pi)^{-1/2}\int_{-\infty}^{\infty} \vec{j}(t) e^{i\omega t} dt.
\end{equation}
Often, a window function (mask) $w(t)$ is multiplied on to the current $\vec{j}(t)$ to smoothly reduce it at large times (mimics current decay from scattering or other decay mechanisms) such that the integral can be done at finite times. For long pulses ($\sim$ ten optical cycles), the qualitative features of the HHG spectrum do not depend on the mask chosen, as long as the emission near the peak of the external field is retained.
The emission intensity for harmonics polarized along a direction $\uv{n}$ is given by
\begin{equation}
  \label{eq:tdse_hhg_3}
  S_{\uv{n}}(\omega) \propto \omega^2\abs{\vec{j}(\omega)\cdot\uv{n}}^2.
\end{equation}
Separate spectra for the decomposed currents can be obtained, for example as $S^\tra(\omega) \propto \omega^2\norm{\vec{j}^\tra(\omega)}^2$. Note however, that generally $S(\omega) \ne S^{\tra}(\omega) + S^\anom(\omega) + S^\mix(\omega) + S^\ter(\omega)$ due to existence of cross terms such as $j_\pol^\tra(\omega)j_\pol^\ter(\omega)$ -- the decomposition of the spectrum only makes sense at those $\omega$ where the cross terms are negligible (in practice, this is almost always the case).

More information on the time-frequency characteristics of the emitted harmonics can be obtained by a continuous wavelet transform (CWT) of the current,
\begin{equation}
  \label{eq:tdse_hhg_4}
  \begin{aligned}
    S(t, a) \propto a^{-2} \norm{\int_{-\infty}^{\infty}
      \vec{j}(t') W^*\left[\frac{t'-t}{a}\right] dt'}^2
  \end{aligned}
\end{equation}
with $t$ a time-translation variable, $a$ a frequency-scaling variable, and $W(x)$ a mother wavelet. In our calculations, we use the Morlet-Grossman mother wavelet, with $W(x)=(\sigma^2\pi)^{-1/4} e^{-i\Omega x}e^{-x^2/(2\sigma^2)}$, where $\sigma$ is the standard deviation, and $\Omega$ the center frequency of the mother wavelet. 
Insertion of $x\equiv (t'-t)/a$ into $W(x)$ shows that the frequency of the daughter wavelet is scaled as $\omega\equiv\Omega/a$, while the time-spread is scaled as $\sigma_{\text{time}}=\sigma a$. Similarly, the daughter wavelet in Fourier space has the spread $\sigma_{\text{freq}}=(\sigma a)^{-1}$.
This scaling means that higher time resolution, along with lower frequency resolution, is employed for the higher frequencies in the HHG spectrum. 
 This can be an advantage over a windowed Fourier transform where the time- and frequency resolution is fixed. The CWT can be efficiently evaluated \cite{Chandre2003PhysicaD} using the convolution theorem on the integral in Eq.~\eqref{eq:tdse_hhg_4}, resulting in
\begin{equation}
  \label{eq:tdse_hhg_5}
  \begin{aligned}
    S(t, a) \propto \norm{ FT^{-1}\left[ \vec{j}(f)W^*(a f) \right] }^2,
  \end{aligned}
\end{equation}
where $W(f)$ denotes the Fourier transform of the mother wavelet. Eq.~\eqref{eq:tdse_hhg_5} is efficient, since $\vec{j}(f)$ is independent of $a$, and $W(f)$ is known analytically. As discussed, the frequency is related to $a$ by $\omega=\Omega/a$, and $a$ is sampled dyadicly.

\begin{figure}
  \centering
  \includegraphics[width=0.7\textwidth, clip, trim=0 0cm 0 0cm]{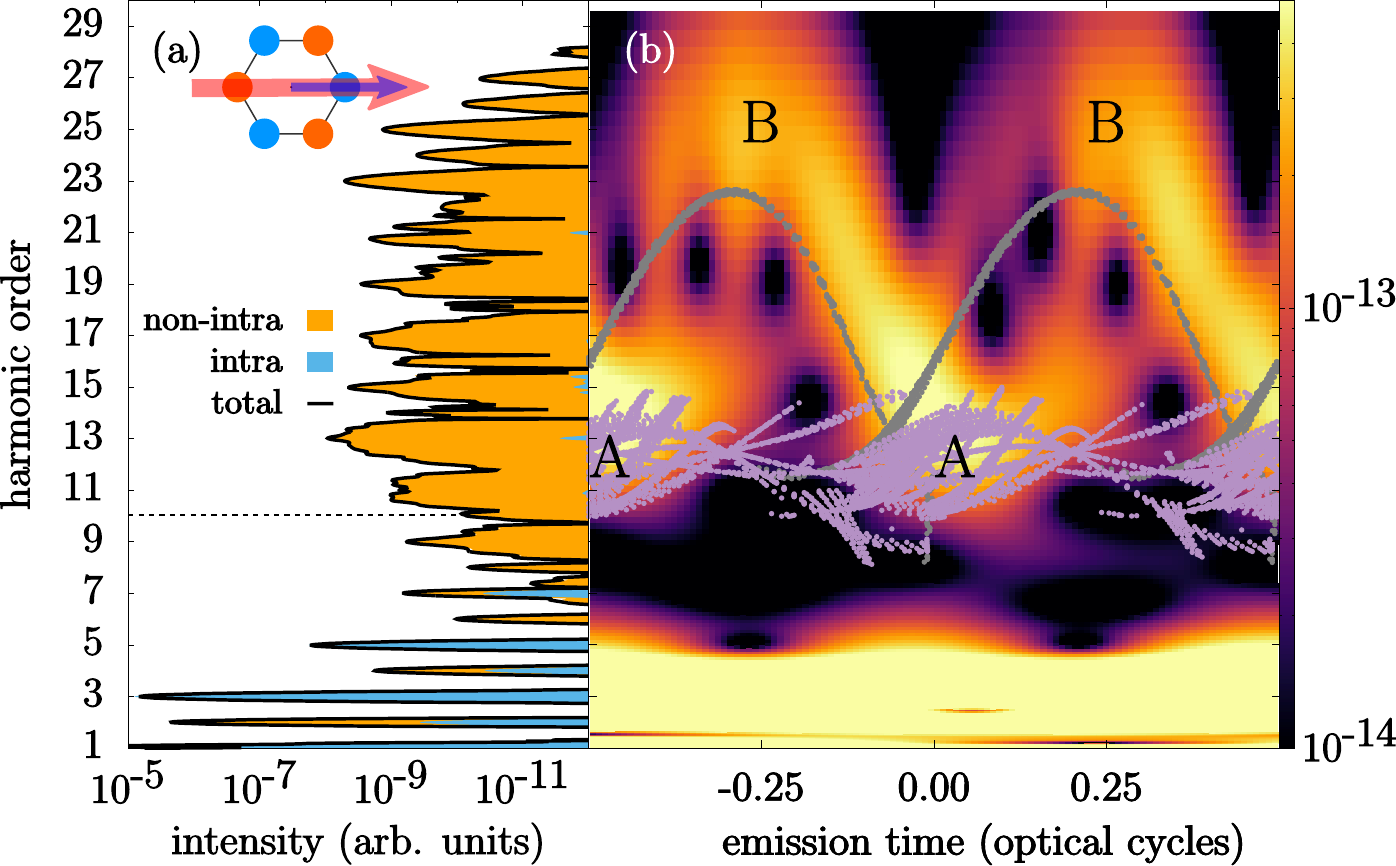}
  \caption{(a) HHG spectrum for \hBN{} irradiated by a 1600 nm, 3.5 TW/cm$^2$, 58.7 fs pulse. The inset shows a sketch of the crystal structure, the driving laser polarization direction (red arrow along the armchair direction) and the detected HHG polarization direction (blue arrow). The horizontal dashed line shows the position of the minimum band gap. (b) Time-frequency profiles for the harmonics. The labels A and B marks two different features in the spectrum. The results from the semiclassical recollision model (see Sec.~\ref{sec:saddle}) are superimposed, with the gray points originating from recollisions of electron-hole pairs created near a $M_1$ symmetry point (a $M$ point located on the line going through $\Gamma$ along the armchair direction), while the purple points are from electron-hole pairs created near the other $M$ points and $K$ points.}
  \label{fig:tdse_hhg_1}
\end{figure}

We now provide a specific example of an HHG calculation. We consider a monolayer of hexagonal boron nitride (\hBN), with the band structure calculated using a pseudopotential method detailed in Ref.~\cite{Taghizadeh2017PRB, Yue2020PRA}. A two-band model is considered, consisting of the highest valence band and the lowest conduction band, with the hexagonal BZ having minimum band gaps at the $K$ high-symmetry points with energy $7.8$ eV. The periodic TPT structure gauge is constructed following the procedure outlined in Sec.~\ref{sec:ham_tpt}, which provides us with smooth and periodic Berry connections and dipole couplings. Prior to the time-propagation, the valence band is assumed fully occupied and the conduction band empty.
We irradiate the monolayer with a laser linearly polarized along the armchair direction $\uv{x}$ [indicated by the red arrow in the inset of Fig.~\ref{fig:tdse_compare_1}(a)], with the vector potential of the form
\begin{equation}
  \label{eq:tdse_hhg_6}
  \vec{A}(t) = A_0 \cos^2\left[ \frac{\pi t}{2\tau}\right]\sin(\omega_0t) \uv{x}, \quad t\in[-\tau, \tau],
\end{equation}
where $A_0=0.35$ ($I=3.5$ TW/cm$^2$), $\omega=0.0285$ ($\lambda=1600$ nm), and $\tau=58.7$ fs.
For the time propagation we employ Eq.~\eqref{eq:tdse_houston_5}, i.e. the SBEs in the moving frame. A Monkhorst-Pack mesh is used with total number of $\vec{K}$-discretization points $300\times 300=9\times 10^4$, and the dephasing time is chosen $T_2=5$ fs.  The HHG spectrum for harmonics polarized along  the arm-chair direction $\uv{x}$ [blue arrow, inset of Fig.~\ref{fig:tdse_hhg_1}(a)] is calculated using \eqref{eq:tdse_hhg_3} with $\uv{n}=\uv{x}$, and shown in Fig.~\eqref{eq:tdse_hhg_1}(a). Harmonics up to the 28th order are observed, consisting of both even- and odd-order harmonics. The spectrum in Fig.~\ref{eq:tdse_hhg_1}(a) is further separated into the intraband and non-intraband contributions from the microscopic current, following Eq.~\eqref{eq:tdse_houston_7}.
Below the band gap energy (dashed line), the odd harmonics are dominated by the intraband contribution, originating from the carrier transport in the individual bands; while above the band gap, the non-intraband contribution dominates, originating from the coupling between the bands; the even-order harmonics below the band gap are due to the anomalous current, which is part of the non-intraband contribution.

The density plot in Fig.~\ref{fig:tdse_hhg_1}(b) shows the time-frequency profiles of the harmonic emissions, obtained using Eq.~\eqref{eq:tdse_hhg_5}. The intraband harmonics below the band gap have a broad time-profile, with the highest-order intraband harmonics [around harmonic 5 (H5)] emitted near $t=n/2$ optical cycles ($n$ an integer), corresponding to the zeros of the vector potential $\vec{A}(t)$ in Eq.~\eqref{eq:tdse_hhg_6}. This can be understood from the dependence of the intraband current on the time-dependent band structure $E_n^{\vec{K}+\vec{A}(t)}$ [Eq.~\eqref{eq:tdse_houston_7a}], where the harmonic emissions occur at times corresponding to the largest band curvature (largest rate of change in the group velocity) \cite{Wu2015PRA}. In the case of \hBN, this is near the $K$ and $M$ high-symmetry points.

The non-intraband harmonics above the band gap energy in Fig.~\ref{fig:tdse_hhg_1}(b) have a distinct emission profile compared to the intraband case. The strongest non-intraband harmonic at H13 [see also Fig.~\ref{fig:tdse_hhg_1}(a)] are emitted at around $t=n/2$ optical cycles, and we have labelled the structure in the time-profiles as ``A''. The time-profiles for the harmonics emitted above 17th order have a bow-like structure, with the highest-order harmonics emitted at around $t_1=(0.25 + n/2)$ optical cycles, which we have labelled ``B''. For $t<t_1$, the slope of the time-profile is positive, corresponding to a positive chirp, while for $t>t_1$ the emissions are negatively chirped. This bow-like structure is similar to the time-frequency profile for HHG in gases, where every energy below the cut-off harmonic is emitted twice, corresponding to the \textit{short} and \textit{long} trajectories \cite{Vampa2017tutorial}. Using shorter dephasing times $T_2$ will suppress the long trajectories and result in more well-resolved HHG spectra. In Sec.~\ref{sec:saddle} we will introduce the semiclassical recollision model for the non-intraband harmonics, and discuss that the structure A and B originate from recollisions of electron-hole pairs created near different symmetry point in the BZ.

\section{Saddle-point equations and the recollision model for HHG} \label{sec:saddle}
In the previous section, we have presented methods to obtain relevant observables such as the microscopic current and the HHG spectrum, by time-propagating the relevant EOM. These black-box calculations can be considered numerical experiments that contain all the relevant information, but they are often too complex to gain physical insights, with everything intermingled. In this section, we discuss methods that can help us gain physical intuition and understanding, especially on the emission dynamics of HHG.

\subsection{Saddle-point method for HHG}
We consider a two-band model with non-degenerate bands that include an initially filled valence band denoted by the band index $v$ and an empty conduction band denoted by $c$. For concreteness, we work in the VG and in the Houston basis from Sec.~\ref{sec:tdse_houston}.
The EOM in Eq.~\eqref{eq:tdse_houston_5} (also referred as the SBEs in the moving frame) reduce to
\begin{subequations}
  \label{eq:saddle_1}
  \begin{align}
    \dot{\bar{\rho}}_{vv}^{\vec{K}}(t) = & i \vec{F}(t)\cdot
                                  \vec{d}^{\vec{K}+\vec{A}(t)}\bar{\rho}_{vc}^{\vec{K}}(t) + \text{c.c.}\\
    \dot{\bar{\rho}}_{cc}^{\vec{K}}(t) = & -i \vec{F}\cdot
                                  \vec{d}^{\vec{K}+\vec{A}(t)}\bar{\rho}_{vc}^{\vec{K}}(t) + \text{c.c.}\\
    \dot{\bar{\rho}}_{cv}^{\vec{K}}(t) = & \left[
                                  - i \omega_g^{\vec{K}+\vec{A}(t)}
                                  - i \vec{F}(t) \cdot  \Delta \vec{\berryc}^{\vec{K}+\vec{A}(t)}
                                  \right] \bar{\rho}_{cv}^{\vec{K}}(t) \nonumber \\
                                   &- i \left[\bar{\rho}_{vv}^{\vec{K}}(t) - \bar{\rho}_{cc}^{\vec{K}}(t)\right] \vec{F}(t) \cdot \vec{d}^{\vec{K}+\vec{A}(t)},
  \end{align}
\end{subequations}
with $\omega_g^{\vec{K}}\equiv E_c^{\vec{K}}-E_v^{\vec{K}}$ the band gap and $\Delta\berryc^{\vec{K}}\equiv\berryc_c^{\vec{K}}-\berryc_v^{\vec{K}} $ the Berry connection difference. For HHG in semiconductors and insulators, the population transfer to the conduction band is small, and we make the approximation $\bar{\rho}_{vv}^{\vec{K}}-\bar{\rho}_{cc}^{\vec{K}}\approx 1$. The formal solutions to Eq.~\eqref{eq:saddle_1} now read
\begin{subequations}
  \label{eq:saddle_2}
  \begin{align}
    \bar{\rho}_{vv}^{\vec{K}}(t) = & i \int^t ds \vec{F}(s) \cdot \vec{d}^{\vec{K}+\vec{A}(s)}\bar{\rho}_{vc}^{\vec{K}}(s) + \text{c.c.}\\
    \bar{\rho}_{cc}^{\vec{K}}(t) = & -i \int^t ds \vec{F}(s) \cdot \vec{d}^{\vec{K}+\vec{A}(s)}\bar{\rho}_{vc}^{\vec{K}}(s) + \text{c.c.}\\
    \bar{\rho}_{cv}^{\vec{K}}(t) = & -i \int^t ds
                               \vec{F}(s) \cdot \vec{d}^{\vec{K}+\vec{A}(s)} e^{- T_2^{-1} (t-s)} \nonumber \\
                             & \times e^{- i\int_s^t
                               \left[ \omega_g^{\vec{K}+\vec{A}(t')} + \vec{F}(t') \cdot \Delta\vec{\berryc}^{\vec{K}+\vec{A}(t')} \right]
                               dt'},
  \end{align}
\end{subequations}
which can easily be checked by insertion.
We are interested in the above-band gap harmonics, which are dominated by the non-intraband contribution, as illustrated by the example shown in Fig.~\ref{fig:tdse_hhg_1}. Insertion of Eq.~\eqref{eq:saddle_2} into Eq.~\eqref{eq:tdse_houston_7b}, and transforming into the fixed frame $\vec{k}\equiv \vec{K}+\vec{A}(t)$ results in
\begin{equation}
  \label{eq:saddle_3}
  \begin{aligned}
    j_\pol^\ntra(t)
    & = N^{-1} \sum_{\vec{k}} R^{\vec{k}}_\pol  \int^t
     T^{\vec{\kappa}(t,s)}
    e^{-iS^\pol(\vec{k}, t, s)} ds
    + \text{c.c.}
  \end{aligned}
\end{equation}
with $\pol=\{x, y, z\}$ the Cartesian indices, $T^{\vec{\kappa}(t,s)}=|\vec{F}(s)\cdot\vec{d}^{\vec{\kappa}(t,s)}|$ the transition matrix element, $R^{\vec{k}}_\pol =  \omega_{g}^{\vec{k}}| {d}_{\pol}^{\vec{k}} |$ the recombination dipole, $\vec{\kappa}(t, t') = \vec{k} - \vec{A}(t) + \vec{A}(t')$ the time-dependent crystal momentum, and c.c. stands for complex conjugate.
The times $s$ and $t$ can be interpreted as the excitation and emission times, respectively.
The accumulated phase in Eq.~\eqref{eq:saddle_3} is
\begin{equation}
  \label{eq:saddle_4}
  \begin{aligned}
    S^\pol(\vec{k}, t, s)
    =& \int_s^t \left[ \omega_g^{\vec{\kappa}(t, t')} + \vec{F}(t')\cdot\Delta\berryc^{\vec{\kappa}(t, t')} \right] dt' \\
    & + \alpha^{\vec{k},\pol}
    - \beta^{\vec{\kappa}(t,s)}
  \end{aligned}
\end{equation}
with $\alpha^{\vec{k},\pol} \equiv\arg({d}_\pol^{\vec{k}})$ the transition-dipole phases \cite{Jiang2018PRL, Li2019PRA}, and $\beta^{\vec{\kappa}(t,s)}\equiv \arg[\vec{F}(s)\cdot \vec{d}^{\vec{\kappa}(t,s)}]$. Note that inclusion of $\alpha^{\vec{k},\pol}$ and $\beta^{\vec{\kappa}(t,s)}$ in $S^\pol(\vec{k},t,s)$ results in it being structure-gauge invariant \cite{Li2019PRA,Yue2021PRA}.

We are interested in the frequency-resolved non-intraband current, $j_\pol^\ntra(\omega) = \int_{-\infty}^{\infty} dt e^{i\omega t} j_\pol^\ntra(t)$, which contributes to the non-intraband HHG spectrum.
The saddle-point approximation in the absence of simple poles consists of only including the stationary (saddle) points of the phase factor involving $S^\pol(\vec{k}, t, s)-\omega t$, since other contributions lead to highly oscillatory terms in the integrand of $j_\pol^\ntra(\omega)$.
Taking the partial derivatives with respect to the three integration variables $\vec{k}$, $s$ and $t$, the saddle point conditions are \cite{Li2019PRA, Yue2021PRA}
\begin{subequations}
  \label{eq:saddle_5}
  \begin{align}
    \omega_g^{\vec{\kappa}(t, s)} + \vec{F}(s) \cdot \ccom^{\vec{\kappa}(t, s)}
    & = 0, \label{eq:saddle_5a}
    \\
    \Delta \vec{R}^\pol \equiv \Delta\vec{r} - \com^{\vec{k},\pol} + \ccom^{\vec{\kappa}(t, s)}
    & = \vec{0}, \label{eq:saddle_5b}
    \\
    \omega_g^{\vec{k}}
      + \vec{F}(t) \cdot \left[ \ccom^{\vec{\kappa}(t, s)} + \Delta\vec{r}  \right]
    & = \omega, \label{eq:saddle_5c}
  \end{align}
\end{subequations}
with the electron-hole separation vector and group velocities
\begin{subequations}
  \label{eq:saddle_6}
  \begin{align}
    \Delta \vec{r} \equiv
    & \int_s^t \left[ \vec{v}_c^{\vec{\kappa}(t, t')} - \vec{v}_v^{\vec{\kappa}(t, t')} \right] dt'   \label{eq:saddle_6a} \\
    \vec{v}_n^{\vec{\kappa}(t, t')} \equiv
    & \nabla_{\vec{k}}E_n^{\vec{\kappa}(t, t')} + \vec{F}(t') \times \vec{\Omega}_n^{\vec{\kappa}(t, t')},   \label{eq:saddle_6b}
  \end{align}
\end{subequations}
and the structure-gauge invariant quantities
\begin{subequations}
  \label{eq:saddle_7}
  \begin{align}
    \com^{\vec{k},\pol} \equiv & \Delta \berryc^{\vec{k}} - \nabla_{\vec{k}}\alpha^{\vec{k},\pol} \label{eq:saddle_7a} \\
    \ccom^{\vec{k}} \equiv & \Delta \berryc^{\vec{k}} - \nabla_{\vec{k}}\beta^{\vec{k}} \label{eq:saddle_7b}.
  \end{align}
\end{subequations}
We denote a saddle point, i.e. a solution to Eqs.~\eqref{eq:saddle_5a}-\eqref{eq:saddle_5c}, as $\{\sad{\vec{k}}, \sad{t}, \sad{s}\}$.
The essense of the interband HHG process is contained in the interpretation of Eqs.~\eqref{eq:saddle_5a}-\eqref{eq:saddle_5c} in terms of the following three steps: an electron-hole pair is created by tunnel excitation at time $\sad{s}$ and with the crystal momentum $\vec{k}_0\equiv \sad{\vec{k}} - \vec{A}(\sad{t}) - \vec{A}(\sad{s})$; the hole and electron are accelerated by the laser with the instantaneous group velocities $v_v^{\sad{\vec{k}}-\vec{A}(\sad{t})+\vec{A}(t')}$ and $v_c^{\sad{\vec{k}}-\vec{A}(\sad{t})+\vec{A}(t')}$, respectively; the electron-hole pair recombine at time $\sad{t}$ with final crystal momentum $\sad{\vec{k}}$ and relative distance $\Delta\vec{r}$, with the simultaneous emission of high-harmonics with energy $\omega$.

The saddle-point conditions cannot generally be satisfied by purely real values of $\vec{k}$, $t$ and $s$, and must generally be solved by analytic continuation of the parameter space into the complex plane.
For example, in band-gap materials where $\mathcal{\com}^{\vec{k},\mu}$ and $\mathcal{\ccom^{\vec{k}}}$ are negligible, Eq.~\eqref{eq:saddle_5a} requires the band gap to be zero, which clearly cannot be satisfied for real values of $\vec{k}$; similarly, Eq.~\eqref{eq:saddle_5b} requires $\Delta \vec{r}=\vec{0}$, which is often too restrictive for real-valued saddle points. Thus a fully rigorous quantum solution would involve complex-valued saddle points and constitute a monumental numerical task, and we mark this as another interesting direction of research for HHG in solids. Recent progress has been made towards approximately solving the saddle-point equations for reduced-dimensionality model systems in Refs.~\cite{Parks2020Optica, Navarrete2019PRA}.
It remains to be seen whether such formalisms can treat electron-hole-pairs starting from different regions in the BZ. 
In the next subsection, we will describe a semiclassical solution to the saddle-points conditions that can provide temporal and spectral insights into the mechanism of interband HHG.

The saddle-point approximation to the frequency-resolved non-intraband current in Eq.~\eqref{eq:saddle_3} involves deforming the integration contours from the real axes into the complex planes, going through the complex saddle points along the path of steepest descent \cite{Arfken2005book} defined by constant $\text{Re}\left[S^\pol(\vec{k},t,s)-\omega t \right]$. The approximation reads in the present case \cite{Uzan2020NPhoton, Navarrete2019PRA}
\begin{equation}
  \label{eq:saddle_8}
  \begin{aligned}
    j_\pol^\ntra(\omega)
    \propto & \sum_{\sad{\vec{k}}, \sad{t}, \sad{s}}
    \frac{R^{\sad{\vec{k}}}_\pol
      T^{\sad{\vec{k}} - \vec{A}(\sad{t})+\vec{A}(\sad{s})}
      e^{-i \left[ S^\pol(\sad{\vec{k}}, \sad{t}, \sad{s}) - \omega \sad{t}\right]} }
      {\sqrt{\det[\partial^2 S^\mu(\sad{\vec{k}},\sad{t},\sad{s})]}}
    + \text{c.c.}(\omega \rightarrow -\omega)
  \end{aligned}
\end{equation}
where we have used the notation $\partial^2 S^\mu$ for the Hessian and the second term denotes complex conjugate of first term and with $\omega\rightarrow -\omega$.
In systems where $\com^{\vec{k},\mu}$ and $\ccom^{\vec{k}}$ can be chosen zero, the Hessian is often proportional to $\|\nabla_{\vec{k}}\omega_g^{\sad{\vec{k}}}\|$ \cite{Uzan2020NPhoton}. This makes it clear  that for small values of $\|\nabla_{\vec{k}}\omega_g^{\sad{\vec{k}}}\|$, i.e. when the valence and conduction bands have similar slopes at the saddle point $\sad{\vec{k}}$, the harmonic yield at the corresponding harmonic energy $\omega=\omega_g^{\sad{\vec{k}}}$ is expected to be greatly enhanced, a phenomenon termed as ``spectral singularities'' in Ref.~\cite{Uzan2020NPhoton}.

\subsection{Semiclassical recollision model}
As mentioned, the full quantum solution to the saddle-point equations in Eq.~\eqref{eq:saddle_5} presents a monumental task, and a solution has only been attempted for very simple model systems. A lot of physical insight and intuition, however, can be gained from a semiclassical solution \cite{Vampa2014PRL, Vampa2015Nature, Vampa2015PRB, Li2019PRA, Yue2020PRL, Yue2021PRA} to the saddle-point equations, which we present below and sketched in Fig.~\ref{fig:intro_1}.

In the semiclassical procedure, an initial tunneling time $s$ and an initial crystal momentum $\vec{k}_0$ is picked, where equation~\eqref{eq:saddle_5a} determines $\vec{k}_0\equiv \vec{\kappa}(t,s)$ at tunneling time $s$. Since this equation generally cannot be satisfied for real-valued $\vec{k}_0$ [the second term of Eq.~\eqref{eq:saddle_5a} is generally small], we choose $\vec{k}_0$ to be at or close to a high-symmetry point with a small band gap, which corresponds to a high tunneling probability. The tunneling time is picked in an optical cycle of the pulse $s\in[-T, 0]$.

The next step in the semiclassical solution involves the propagation of the integral in Eq.~\eqref{eq:saddle_6a}, i.e. calculating the electron and hole classical real-space motions for $t'\in[s,t]$ with the time-dependent crystal momentum $\vec{\kappa}(t,t') = \vec{k}_0+\vec{A}(t')-\vec{A}(s)$ and the group velocities given in Eq.~\eqref{eq:saddle_6b}. Equation~\eqref{eq:saddle_5b} denotes the electron-hole recollision condition, i.e. when the electron-hole distance $\norm{\Delta \vec{r}}$ is equal to $\norm{\com^{\vec{k},\pol} - \ccom^{\vec{k}_0}}$, where the latter quantity is generally small in the systems considered by us. For real-valued saddle points, this condition is often too restrictive due to the complicated band dispersions in a crystal - in contrast to the free-electron dispersion relevant for HHG in gases. We thus relax the recollision condition: at each $t'$ during the time-propagation, we calculate $\Delta \vec{R}^\pol$, and record a semiclassical recollision event if (i) $\norm{\Delta \vec{R}^\pol}$ as a function of $t'$ is a local minimum and (ii) $\norm{\Delta\vec{R}^\pol}<R_0$ is fulfilled, with $R_0$ a preset recollision threshold value. The threshold $R_0$ is chosen as the minimum $R_0$ such that the recollision-model results agree with the time-profiles obtained from the quantum simulation results. Often, $R_0$ can be multiple times greater than the lattice constants of the crystal.
The last saddle-point equation~\eqref{eq:saddle_5c} determines the electron-hole recollision energy $\omega$, and thus the energy of the emitted harmonics.

As discussed above, the relaxation of the electron-hole recollision condition means that we allow for \textit{imperfect recollisions} where $\norm{\Delta \vec{R}^\mu}\ne \vec{0}$ \cite{Crosse2014NCommun, Zhang2019PRB, Yue2020PRL, Yue2021PRA}. One consequence of such an recollision event is that the harmonic energy $\omega$ attains an extra electron-hole-pair polarization energy $\vec{F}(\sad{t})\cdot \Delta\vec{r}$  \cite{Yue2020PRL}. Physically, this energy constitutes the potential energy of the electric dipole comprised of the electron-hole pair at the time of recollision. Clearly, imperfect recollisions will occur whenever the different Cartesian components of $\Delta\vec{R}^\pol$ are nonzero or zero at different times, i.e. whenever the direction of motion of the time-dependent crystal momentum $\partial_{t'}\vec{\kappa}(t,t') =-\vec{F}(t')$ is not along the instantaneous group velocities $\vec{v}_n^{\vec{\kappa}(t,t')}$. This condition can e.g. be satisfied for systems driven by elliptically polarized pulses or in systems with large Berry curvatures.
The nonzero recollision distance is a consequence of the spatially extended electron and hole wavepackets, which in periodic systems can span over several unit cells \cite{Crosse2014NCommun, Kruchinin2018review, Yue2020PRL}. At the time of recollision, even if the electron-hole centers are displaced, the wave packets can still overlap and ensure a recollision event. Indeed, the minimum value of the chosen recollision threshold $R_0$ at which the semiclassical recollision model and quantum emission profiles agree is a qualitative measure of the width of the wave packets.
Lastly, even though the presented semiclassical model assumes the birth of the electron-hole pair with zero spatial dispacement, recent work \cite{Parks2020Optica, Yu2021arxiv} has indicated that the electron and hole can emerge dispaced in real-space after tunneling.

In the monolayer-\hBN{}-HHG example presented in Sec.~\ref{sec:tdse_hhg}, the semiclassical result for the recollision energy versus recollision time is plotted in Fig.~\ref{fig:tdse_hhg_1}(b) as points superimposed on the time-frequency profiles. The recollision events represented by the dark gray points originate from electron-hole pairs initially created in a disk of radius $\Delta k=0.1$ around a $M_1$ symmetry points\footnote{A $M_1$ point is defined as a $M$ point located on the line going through the $\Gamma$ point along the armchair direction.}, while the purple points originate from electron-hole pairs initially created near the other $M$ points and $K$ points.
The agreement of the semiclassical result with the time-frequency profile is evident, and more physical understanding is gained as the semiclassical model is able to attribute the different structures in the density plot (denoted as structure A and B in Sec.~\ref{sec:tdse_hhg}) to different tunneling sites in reciprocal space. Since many regions in the BZ are relevant for HHG in solids, the time-frequency profiles and subcycle emission dynamics can be quite complicated for different materials and laser polarizations \cite{Vampa2015PRB, Zhang2019PRB, Uzan2020NPhoton, Yu2021arxiv, Yue2021PRA}. At the same time, however, the fact that electron-hole pairs created in different BZ regions leads to distinct harmonic time-frequency characteristics can potentially facilitate the all-optical reconstruction of the band structure in the whole BZ, and not only near the minimum band gap as demonstrated in Ref.~\cite{Vampa2015PRL}.

Note that a natural procedure to obtain a full solution to the saddle point equations \eqref{eq:saddle_5} would be first to first pick a harmonic frequency $\omega$, and then find the desirable saddle points by some numerical procedure. In contrast, in the approximate semiclassical solution, we pick an initial tunneling time $s$ and crystal momentum $\vec{k}_0$, and check the above-described recollision conditions on-the-fly during the classical time-propagation of a trajectory. If the recollision conditions are satisfied, we record it as a recollision event and say that we have found a saddle point solution $\{\sad{\vec{k}},\sad{t},\sad{s}\}$. Due to the deterministic nature of the classical trajectories, it is numerically manageable to pick all combinations of the tunneling time in an optical cycle $s\in[-T, 0]$ and the crystal momentum $\vec{k}_0$ in a sphere around the high-symmetry points, as well as propagate the trajectories up to two optical cycles after tunneling $t\in [s,s+2T]$.

\section{Macroscopic propagation} \label{sec:macro}
Up to this point, we have mostly discussed ways to treat the microscopic dynamics responsible for HHG in solids. However, when electromagnetic radiation propagates through a medium, the medium responds by not only generating new radiation as discussed above, but sometimes also by modifying the propagating radiation.
In the field-intensity regime of pulsed lasers, the propagation of the electromagnetic fields can be described classically, i.e. governed by Maxwell's equations
\begin{subequations}
  \label{eq:macro_1}
  \begin{align}
    \nabla \cdot \vec{F} = & \epsilon_0^{-1}\rho  \label{eq:macro_1a} \\
    \nabla \cdot \vec{B} = & 0  \label{eq:macro_1b} \\
    \nabla \times \vec{F} = & -\partial_t\vec{B}  \label{eq:macro_1c} \\
    \nabla \times \vec{B} = & \mu_0 \left(\vec{J} +\epsilon_0 \partial_t\vec{F} \right)  \label{eq:macro_1d},
  \end{align}
\end{subequations}
with $\epsilon_0$ the vacuum permittivity, $\mu_0$ the vacuum permeability, $\rho$ the (induced) charge density, $\vec{J}$ the current density proportional to the microscopic current $\vec{j}$ from Sec.~\ref{sec:tdse}\footnote{We thus consider here the case where there are no free charges or currents in the medium.}.
Generally, all sources and fields in Eq.~\eqref{eq:macro_1} are dependent on space $\vec{r}$ and time $t$, which we have omitted for notational clarity.
For small intensities, the medium response is linear with respect to the fields, and governed by the linear constitutive equations involving linear susceptibilities. At higher intensities, the nonlinear response can be modelled by a series expansion in terms of the nonlinear susceptibilities. When such a series expansion fails, as is often the case for the intensity regimes responsible for the highly nonlinear recollision mechanisms and the non-intraband currents, one has to resort to numerical solutions of the microscopic response, which was the topic in the previous sections.

In terms of the vector potential $\vec{A}$ and scalar potential $\Phi$ defined in Eq.~\eqref{eq:tdse_2}, the inhomogeneous wave equations can easily be obtained from Eq.~\eqref{eq:macro_1a} and Eq.~\eqref{eq:macro_1c} by taking the curl of $\vec{B}$,
\begin{equation}
  \label{eq:macro_2}
  \begin{aligned}
      \nabla^2\Phi + \partial_t\left(\nabla \cdot \vec{A}\right) &= -\epsilon_0^{-1} \rho \\
      c^{-2} \partial_t^2\vec{A} - \nabla^2\vec{A} + \nabla(c^{-2} \partial_t\Phi + \nabla\cdot \vec{A}) & = \mu_0 \vec{J}.
  \end{aligned}
\end{equation}
In principle, the source terms can be obtained from the microscopic calculations at each time and position, and the final coupled Maxwell-Schr\"odinger equations should then be solved. However, due to the insurmountable computational complexities involved, appropriate approximations are usually required. One often-used approximation is to define two spatial scales with two numerical grids: one macroscopic scale with coordinate $\vec{R}$ on the order of the electromagnetic wavelength $\lambda$ to treat the pulse propagation, and a smaller scale $\vec{r}$ on the order of unit cells ($\ll \lambda$) where a local dipole approximation for the microscopic physics can be made. One advantage of using the scalar and vector potentials instead of the physical fields is the flexibility of choosing different laser gauges in the two different scales. E.g. one of the gauges in Sec.~\ref{sec:tdse} can be used for the microscopic physics, while a gauge with $\Phi \equiv 0$ can be used for the macroscopic scale \cite{Yabana2012PRB}.
To further reduce the problem, an approximate one-dimensional propagation scheme can be used, where
the laser pulse propagates along $Z$ (perpendicular to a crystal surface), such that the vector potential in the local dipole approximation reads $\vec{A}_{\vec{R}}(t)=A_Z(t)\uv{e}$, with $\uv{e}$ the laser polarization direction. The symmetry of the setup is assumed to be such that the generated current has the same polarization as the laser, and $\vec{J}_{\vec{R}}(t)=J_Z(t)\uv{e}$. The wave equation reduces in this case to
\begin{equation}
  \label{eq:macro_3}
  c^{-2} \partial_t^2A_Z - \partial_Z^2A_Z  = \mu_0 J_Z,
\end{equation}
with $J_Z$ the current obtained from the microscopic calculation.
Such an approach has e.g. been used to show \cite{Floss2018PRA} that the HHG spectrum after propagation through a bulk crystal exhibits a much ``cleaner'' spectrum (more well-resolved harmonic peaks) compared to the purely microscopic result, which is attributed to the destructive interference between electron-hole recollision events at different recombination times along the propagation path. Very recently, the optimal thickness for HHG in silicon thin films has been investigated \cite{Yamada2021PRB}.
Note however, that this scheme does not include the radial variation of the laser and thus does not include many nonlinear optical effects such as self-focusing.
In addition, the second-order derivative of the propagation coordinate $Z$ in Eq.~\eqref{eq:macro_3} requires a dense discretization along $Z$, with the maximum realizable propagation distance a couple of micrometers \cite{Yabana2012PRB, Floss2018PRA, Yamada2021PRB}.

A different approach is to consider the wave equation for the physical fields
\begin{subequations}
  \label{eq:macro_4}
  \begin{align}
    c^{-2} \partial_t^2{\vec{F}} - \nabla^2 \vec{F} &= -\epsilon_0^{-1}\nabla\rho - \mu_0\partial_t\vec{J}  \\
    c^{-2} \partial_t^2{\vec{B}} - \nabla^2 \vec{B} &= \mu_0(\nabla\times \vec{J}).
  \end{align}
\end{subequations}
For electric fields slowly varying in the transverse dimensions, the scalar approximation for the wave equation \eqref{eq:macro_4} can be made \cite{Brabec2000review}.
 For nonrelativistic carriers, the magnetic field $\vec{B}$ can be neglected, and the wave equation reduces to 
\begin{subequations}
  \label{eq:macro_5}
  \begin{align}
    c^{-2} \partial_t^2F - \nabla^2 F &= - \mu_0\partial_tJ
  \end{align}
\end{subequations}
where the right-hand side of Eq.~\eqref{eq:macro_5} again represents the source terms generated by the medium in response to the propagating field. 
This equation has a second-order derivative along the propagation direction $z$, which requires significant numerical effort.
The computational effort can be significantly reduced by first transforming to a frame that moves at the speed of light, and then ignoring the second derivative with respect to the new $z$ compared to $\frac{2}{c} \partial_{t}\partial_{z}$ and the transverse derivative:
\begin{equation}
  \label{eq:macro_6}
  \begin{aligned}
    \left[ \nabla_\perp^2
      - \frac{2}{c} \partial_{t}\partial_{z}
    \right] F  = \mu_0 \partial_{t} J_0.
  \end{aligned}
\end{equation}
This approximation is usually termed the slowly evolving wave approximation \cite{Brabec1997PRL, Brabec2000review, Kolesik2002PRL, Boyd2008book, Gaarde2011PRA} and also implies ignoring backward-propagating waves. Eq.~\eqref{eq:macro_6} can be conveniently solved in the spectral domain.

Many similar variations of the envelope propagation Eq.~\eqref{eq:macro_6} exist in the literature, where the current or polarization often is separated into linear and nonlinear parts $J=J^{(1)}+J^{NL}$ and the linear part is expressed in terms of the linear susceptibilities and permittivities.
One example is the unidirectional pulse propagation equation (UPPE), which assumes that the nonlinear response is purely due to the forward propagating wave \cite{Kolesik2002PRL, Kolesik2004PRE}. Recently, the UPPE was applied to HHG in solids to show that for propagation lengths longer than the laser wavelength, the propagation significantly reduces the HHG yield and can potentially be responsible for the short dephasing times $T_2$ used in microscopic simulations to match with experiments \cite{Kilen2020PRL}.
The UPPE has also in its full vectorial form been used to show that the term in Eq.~\eqref{eq:macro_4} involving $\epsilon_0^{-1}\rho=\nabla\cdot\vec{F}$ in some cases
% $\nabla\cdot\vec{F}$ 
can have a pronounced effect on the self-focusing of ultrashort pulses \cite{Kolesik2002PRL}.

\subsection{Spatio-spectral properties of solid-state HHG}
\begin{figure}
  \centering
  \includegraphics[width=0.5\textwidth, clip, trim=0 0cm 0 0cm]{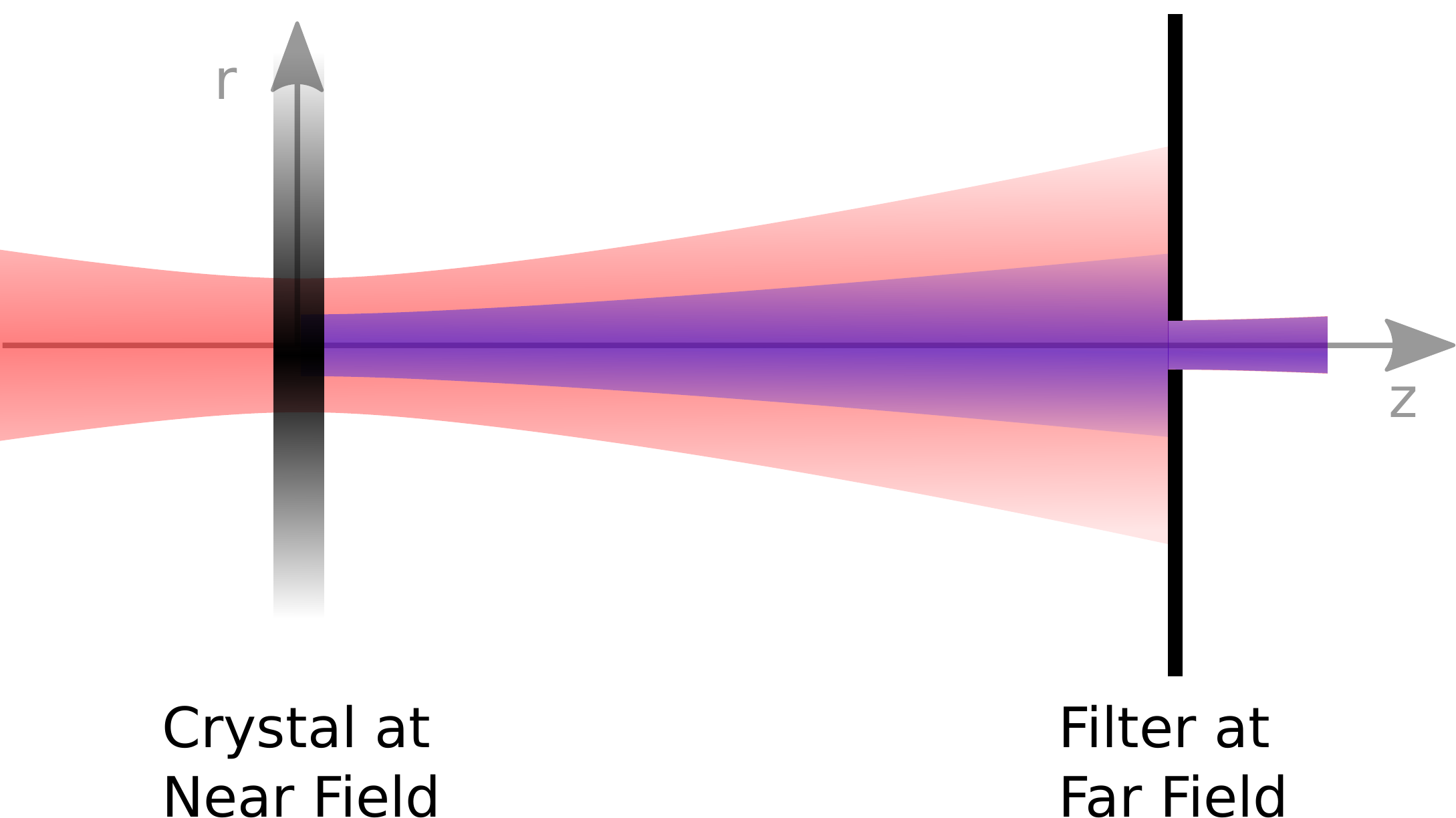}
  \caption{Sketch showing a numerical or (potentially) experimental setup to probe the spatio-spectral properties of solid-state HHG in which an incoming laser beam hits a thin sample and produces HHG in the near field. The fields are then propagated from the near field to the far field, where a spatial filter for example can select the on-axis radiation, essentially filtering out contributions from undesired electron-hole recollisions in the near-field generation process (see text).}
  \label{fig:macro_1}
\end{figure}

In this section, we give an example of a spatio-spectral analysis of solid-state HHG. The conceptual sketch of our numerical setup is shown in Fig.~\ref{fig:macro_1}: an incoming laser beam propagates normal to a very thin crystal sample along the $z$ axis, hits the sample and generates high-order harmonics; the beam, together with the generated harmonics propagate from the sample (\textit{near field}) toward an opaque screen located a great distance away from the sample (\textit{far field}); the opaque screen has a circular aparture that filters away the parts of the beam that has a large spatial divergence; Optionally, beyond the circular aparture, a focusing lens can be installed to focus the filtered spectrum back to conditions that mimics the near field. The spatially resolved HHG spectrum and time-profiles often provides a separation of different contributions to the HHG process as we will discuss below.

\begin{figure}
  \centering
  \includegraphics[width=0.6\textwidth, clip, trim=0 0cm 0 0cm]{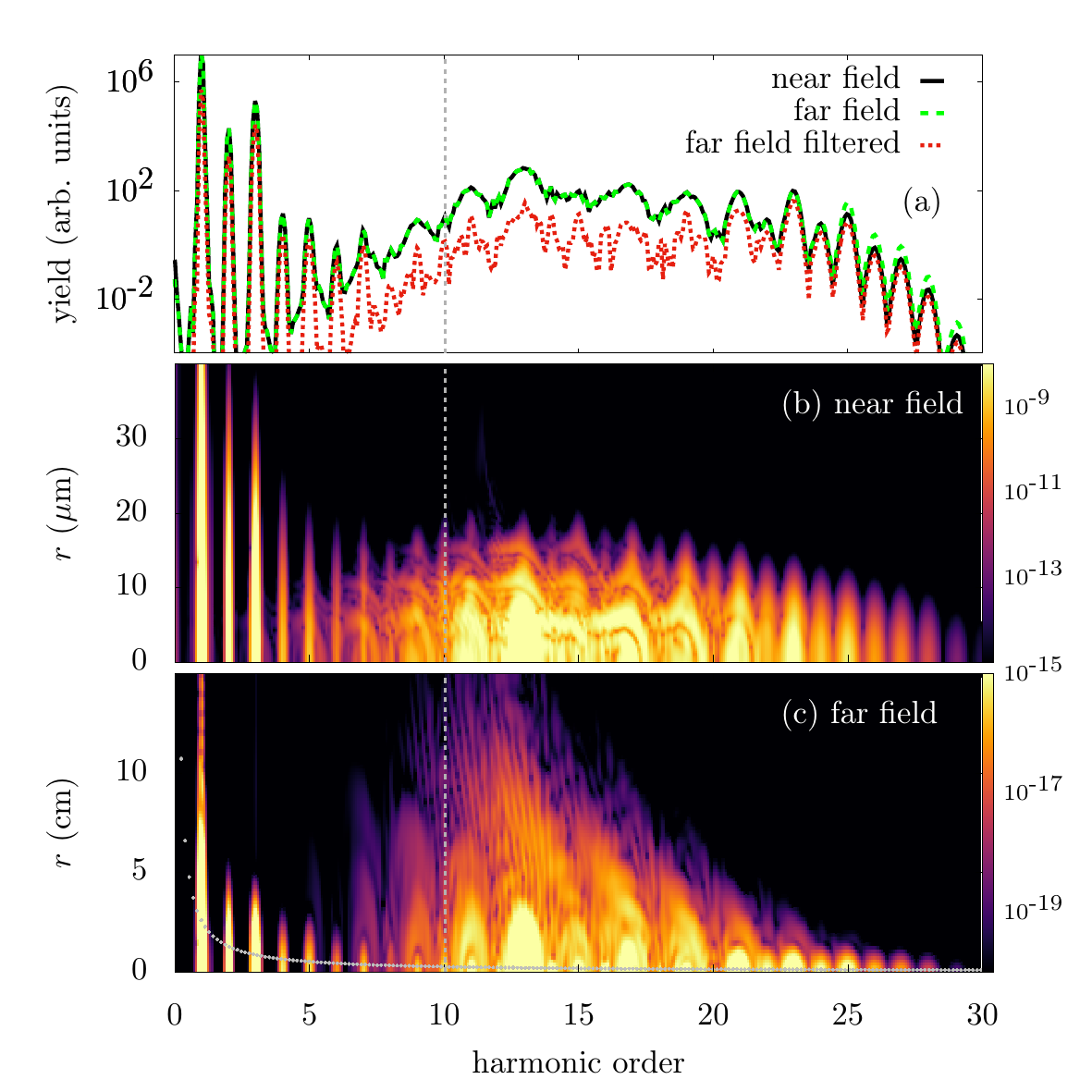}
  \caption{Near- and far-field HHG spectra, for a monolayer of \hBN{} irradiated by a Gaussian beam with waist $w_0=20$ $\mu$m and pulse parameters 1600 nm, 3.5 TW/cm$^2$, 58.7 fs. (a) Near-field, far-field and filtered far-field radially-integrated HHG spectra. (b) Radially-resolved near-field HHG spectrum. (c) Radially-resolved far-field HHG spectrum. The vertical dashed line marks the minimum band-gap. The gray dotted curve plots the divergence radius in the far field expected for Gaussian beams with different carrier frequencies.}
  \label{fig:macro_2}
\end{figure}

For the incident beam, we assume a Gaussian beam $\vec{F}(\vec{r}) = F_0(r,z)e^{ikz} \uv{e}$
\begin{equation}
  \label{eq:macro_ex_1}
  F_0(r,z) = C\frac{w_0}{w(z)} e^{-\frac{r^2}{w(z)^2}} e^{i\left[\frac{kr^2}{2R(z)}-\tan^{-1}\left(\frac{z}{z_0}\right)\right]},
\end{equation}
where $z$ is the propagation direction, $r$ the radial coordinate measured from the beam axis, $k=2\pi n/\lambda$ the wave number, $w(z)=w_0\sqrt{1+z^2/z_0^2}$ the beam waist, $z_0=\pi w_0^2n/\lambda$ the Rayleigh range, $R(z)=z+z_0^2/z$ the radius of curvature, and $C$ a constant determining the pulse amplitude.

We take the crystal sample to be a monolayer of \hBN{} (see Sec.~\ref{sec:tdse_hhg}), placed at the Gaussian beam focus $z=0$. The pulse parameters are chosen as the same as in Sec.~\ref{sec:tdse_hhg}, i.e. 1600 nm, 3.5 TW/cm$^2$ peak intensity and 58.7 fs pulse duration. At the sample, the radial intensity profile of the beam is $I_0(r,0) = \abs{C}^2e^{-2r^2/w_0^2}$, with $w_0=20$ $\mu$m and we use it to calculate the microscopic response at each $r$, resulting in the frequency- and radial-dependent near-field $F_{near}(\omega,r)\propto\omega J(\omega,r)$. The microscopic dynamics is solved using the VG EOM covered in Sec.~\ref{sec:tdse_vg} (no dephasing), including one valence and nine conduction bands, and the pulse parameters are chosen same as in Fig.~\ref{fig:tdse_hhg_1}. The HHG spectrum, $\abs{F_{near}(\omega,r)}^2$, is plotted in Fig.~\ref{fig:macro_2}(b). For harmonics in the interval H10-H20, the spectrum shows complicated behavior compared to the interval H20-H30, reflecting the fact that the former interval has contributions from electron-hole recombinations originating from different BZ symmetry points, as well as the interference of the short and long trajectories [see Fig.~\ref{fig:tdse_hhg_1} and accompanying main text].
The total near-field spectrum is given by $S_{near}(\omega)\propto \int_0^{\infty} \abs{F_{near}(\omega,r)}^2rdr$, which is shown in Fig.~\ref{fig:macro_2}(a) by the black curve. Again, all harmonics in the interval H20-H30 are well-resolved, while only some of the odd harmonics in the interval H10-20 are visible.
Note that since we are dealing with a monolayer material, the macroscopic propagation in the sample is not required. 

In an actual experiment, the photodetector is placed far from the sample, and thus the spectra are only detected in the far-field. As we will see, the far-field spectrum carries information on the underlying HHG process at the near-field. For a Gaussian beam profile known at $z=0$, using the paraxial wave equation, the beam profile at $z>0$ is given as a Hankel transform \cite{Milonni2010book, Peatross2010book, Abadie2018OL} %Peatrossbook10.27
\begin{equation}
  \label{eq:macro_ex_2}
  \bar{F}(\omega,r,z) = -\frac{ik e^{i\frac{kr^2}{2z}}e^{ikz}}{z}\int_0^{\infty} F_{near}(\omega,r) e^{i\frac{kr'^2}{2z}} J_0\left(\frac{krr'}{z}\right) r'dr', 
\end{equation}
with $J_0$ the zeroth order Bessel function and the integration is performed over the near-field radial coordinate $r'$. Equation~\eqref{eq:macro_ex_2} is the same formula as for the diffraction from a circular aparture in the Fresnel approximation.
We calculate the far-field spectrum as $F_{far}(\omega,r)\equiv \bar{F}(\omega,r,z=L)$, with $L=1$m. The total integrated spectrum is accordingly $S_{far}(\omega)\propto \int_0^{\infty} \abs{F_{far}(\omega, r)}^2rdr$.

Figure~\ref{fig:macro_2}(c) shows the $r$-dependent far-field spectrum $|F_{far}(\omega,r)|^2$. 
As observed by comparing the $r$-axis between Figs.~\ref{fig:macro_2}(a) and \ref{fig:macro_2}(b), the beam has diverged appreciably. For the low-order harmonics H1-H9, the divergence decreases with increasing harmonic order, following the divergence angle for a Gaussian beam $\theta=2c/(n\omega w_0)$ \cite{Milonni2010book}, %MilonniEberlybook13.9.1
which is plotted as the gray dotted line. Starting at around the band gap energy ($\sim$H10), the divergence angle of the harmonic radiation has a local maximum, and then decreases again for larger frequencies. This behavior indicates that the harmonics below and above the bandgap are dominated by two distinct microscopic HHG mechanisms. This is in agreement with our discussion in Sec.~\ref{sec:tdse_hhg}, which showed that the below-band gap harmonics are dominated by the intraband current, while the above-bandgap harmonics originate primarily from the non-intraband current. The spatiotemporal profile of the HHG clearly encodes this information. Due to energy conservation, the total far-field harmonic spectrum $S_{far}(\omega)$ is identical to the near-field spectrum $S_{near}(\omega)$, as shown by comparing the green dashed line and the black line in Fig.~\ref{fig:macro_2}(a).

For the interband harmonics, the radial divergence is related to the radial dependence of the accumulated phase of an electron-hole pair between tunneling time $\sad{s}$ and recombination time $\sad{t}$ in the near field, i.e. $S^\mu(\sad{\vec{k}},\sad{t},\sad{s})$ from Eqs.~\eqref{eq:saddle_3} and \eqref{eq:saddle_4}. The steeper the radial phase function (defined as the accumulated phase as a function of $r$) for a harmonic, the larger its radial profile becomes in the far field \cite{Abadie2018OL}. Consequently, the harmonics detected near the beam axis $r=0$ in the far field mostly originate from recollided electron-hole pairs in the near field with small travel times between tunneling and recollision. We can isolate this near-axis spectrum and related trajectories by placing a filter of radius 1 cm in the far field (see Fig.~\ref{fig:macro_1}), and the resulting spectrum $S_{filter}(\omega)\propto \int_0^{r_{filter}} \abs{F_{far}(r, \omega)}^2rdr$ is shown in Fig.~\ref{fig:macro_2}(a) as the red dotted line. In the filtered spectrum, more harmonics are discernible, e.g. H14 and H16 have become visible.

Clearly, the addition of the radial degree of freedom provides additional understanding of the underlying recollision dynamics of solid-state HHG, and can also potentially provide realistic experimental pathways to probe the HHG process. For example, a far-field spectrum was experimentally measured in Ref.~\cite{Luu2015Nature} to confirm the spatial coherence of the generated extreme ultraviolet high-harmonic radiation in SiO$_2$. In Ref.~\cite{Lu2019NPhoton}, interferometry of the dipole phase in HHG was performed by experimentally measuring the far-field spectrum of two overlapping beams in the near-field.

\section{Summary and outlook}
In this tutorial, we have given a hands-on introduction to the theory of HHG in solids. In Sec.~\ref{sec:berry}, we discussed the adiabatic states, the Berry phase and related concepts. In Sec.~\ref{sec:ham}, we described the time-independent problem of a crystalline solid and methods for structure gauge constructions for both nondegerate and degenerate cases. In Sec.~\ref{sec:tdse}, we covered approaches to describe the microscopic HHG mechanism in different laser gauges and structure gauges, and pointed out the advantages and drawbacks for the different methods, as well as provide an example HHG analysis for a monolayer material. In Sec.~\ref{sec:saddle} we discussed the saddle-point approximation to HHG. The semiclassical solutions to the saddle point equations, also termed the recollision model, could reveal spectro-temporal information about the HHG process. In Sec.~\ref{sec:macro}, we formulated ways to describe the macroscopic HHG process, which involved the coupling of the microscopic dynamics to Maxwell's equations. We provided an example of  a monolayer irradiated by a Gaussian beam and discussed the spatio-spectral properties of the HHG detected in the far-field.

Solid-state HHG is a rapidly expanding field, and many emerging theory trends are emerging that are beyond the scope of this tutorial. These include HHG in topological insulators \cite{Bauer2018PRL, Juerss2020PRA, Chacon2020PRB}, doped and amorphous systems \cite{Huang2017, Almalki2018}, and strongly correlated systems \cite{Silva2018NPhoton, Murakami2021PRB, Orthodoxou2021npjQM}. We hope that this work will allow newcomers to get an overview of the topic, as well as provide the necessary tools to perform simulations and stimulte the development of new theory themselves.

\begin{backmatter}
  \bmsection{Funding}
  National Science Foundation (PHY-1713671 and PHY-2110317). Air Force Office of Scientific Research (FA9550-16-1-0013; supported development in Section \ref{sec:tdse_houston}).

  \bmsection{Acknowledgments}
  Portions of this research were conducted with high performance computational resources provided by the Louisiana Optical Network Infrastructure (http://www.loni.org). LY thanks Shicheng Jiang and Francois Mauger for useful discussions.

  \bmsection{Disclosures}
  The authors declare no conflicts of interest.

  \bmsection{Data Availability Statement}
  Data underlying the results presented in this paper are not publicly available at this time but may be obtained from the authors upon reasonable request.
%   \textcolor{red}{No data were generated or analyzed in the presented research.}
  
%  \bmsection{Supplemental document}
%  See Supplement 1 for supporting content. 

\end{backmatter}

\bibliography{ref_proc_solid,ref_books,ref_struc_solid,ref_laser_pulses,ref_manybody_methods,ref_atto_science,ref_strong_field_theories,ref_suppmat,ref_lun,ref_math}

\end{document}